\documentclass{aastex63}

\accepted{May 11, 2020}

\submitjournal{ApJ}

\shorttitle{HXR emission during flux rope eruption}
\shortauthors{Sahu et al.}

\begin{document}

\title{HXR emission from an activated flux rope and subsequent evolution of an eruptive long duration solar flare}

\correspondingauthor{Suraj Sahu}
\email{suraj@prl.res.in}

\author{Suraj Sahu}
\affiliation{Udaipur Solar Observatory, Physical Research Laboratory, Udaipur 313001, India}
\affiliation{Indian Institute of Technology Gandhinagar, Gujarat 382355, India}

\author{Bhuwan joshi}
\affiliation{Udaipur Solar Observatory, Physical Research Laboratory, Udaipur 313001, India}

\author{Prabir K. Mitra}
\affiliation{Udaipur Solar Observatory, Physical Research Laboratory, Udaipur 313001, India}

\author{Astrid M. Veronig}
\affiliation{Institute of Physics, University of Graz, A--8010 Graz, Austria}
\affiliation{Kanzelh{\"o}he Observatory for Solar and Environmental Research, University of Graz, A--9521 Treffen, Austria}

\author{V. Yurchyshyn}
\affiliation{Big Bear Solar Observatory, New Jersey Institute of Technology, Big Bear City, CA 92314, USA}

\begin{abstract}
\noindent In this paper, we present a comprehensive study of the evolutionary phases of a major M6.6 long duration event (LDE) with special emphasize on its pre-flare phase. The event occurred in NOAA 12371 on 2015 June 22. A remarkable aspect of the event was an active pre-flare phase lasting for about an hour during which a hot EUV coronal channel was in build-up stage and displayed co-spatial hard X-ray (HXR) emission up to energies of 25 keV. As such, this is the first evidence of HXR coronal channel. The coronal magnetic field configuration based on NLFFF modeling clearly exhibited a magnetic flux rope (MFR) oriented along the polarity inversion line (PIL) and co-spatial with the coronal channel. We observed significant changes in the AR's photospheric magnetic field during an extended period of $\approx$42 hours in the form of rotation of sunspots, moving magnetic features, and flux cancellation along the PIL. Prior to the flare onset, the MFR underwent a slow rise phase ($\approx$14 km s$^{-1}$) for $\approx$12 min which we attribute to the faster build-up and activation of the MFR by tether-cutting reconnection occurring at multiple locations along the MFR itself. The sudden transition in the kinematic evolution of the MFR from the phase of slow to fast rise ($\approx$109 km s$^{-1}$ with acceleration $\approx$110 m s$^{-2}$) precisely divides the pre-flare and impulsive phase of the flare, which points toward the feedback process between the early dynamics of the eruption and the strength of the flare magnetic reconnection.

\end{abstract}



\section{Introduction} \label{sec:intro}
Solar eruptions are complex phenomena with multiple facets right from their genesis in the solar atmosphere to subsequent consequences in the near-Sun, interplanetary, and near-Earth regions \citep{Gopalswamy2001,Webb2012,
Archontis2019}. Decades of observational and theoretical research has elucidated different aspects of it, namely, solar flares, eruptive prominences, coronal mass ejections (CMEs), coronal jets, etc., which are observationally defined as disjoint terms but occur as a result of physically coupled processes \citep[e.g., see reviews by][]{Priest2002,Fletcher2011}. Exploration of solar eruptive phenomena using multi-wavelength, multi-instrument, and multi-point observations is key toward better understanding of the origin and prediction of space weather events \citep{Koskinen2017,Green2018}.

The source regions of solar eruptions frequently show the presence of interesting observational features, e.g., prominences, filament channels, hot coronal channels, etc., which have been accepted as evidences of a fundamental structure called magnetic flux rope (MFR) \citep{Cheng2011,Patsourakos2013,Joshi2017,Mitra2018,Veronig2018}. MFRs are often defined as a bundle of magnetic field lines that are twisted around each other and wrap around a common axis \citep{Gibson2006, Canou2010, Filippov2015, Cheng2017}. MFRs not only play a crucial role in triggering the eruption but also constitute a key component of CMEs. Near-Earth in-situ measurements often reveal evidence of MFRs at large-scales in the form of interplanetary magnetic clouds signifying the arrival of Earth-directed CMEs \citep{Burlaga1981,Klein1982,
Burlaga1998,Mostl2009,Syed2019}, that may subsequently cause geomagnetic disturbances when interacting with Earth's magnetic field \citep{Zhang1988,Burlaga2001, Zurbuchen2006, Bisoi2016,Joshi2018}. Here some basic questions arise: how do MFRs originate in the solar corona and what are the mechanisms responsible for their eruptions? The analysis of multi-wavelength solar observations of the source regions of CMEs and their comparison with coronal magnetic field modeling yield important insights on these open issues.

The formation of a CME requires the activation and successful eruption of the MFR against solar gravity and the overlying coronal magnetic field. According to the ``standard flare model" also known as CSHKP model \citep{Carmichael1964, Sturrock1966, Hirayama1974, Kopp1976}, the eruptive expansion of the unstable MFR creates strong inflow of plasma and magnetic field lines in the large-scale current sheet that is formed underneath it causing the onset of magnetic reconnection. During magnetic reconnection, the stored magnetic energy is released in the form of intense heating within a localized region as well as acceleration of plasma and high-energy particles \citep{Priest2002,Holman2011}. The spatio-temporal characteristics of a solar flare explored from multi-wavelength and multi-band measurements provide useful information about the origin of the non-thermal and thermal emissions \citep{Fletcher2011, Benz2017}. These observations also pose constraints on the standard flare model \citep{Sui2004,Veronig2004,
Joshi2012}.

It has been observed that many flares are associated with pre-flare and precursor activities, which include small-scale brightness enhancements in the flaring region of about a few to tens of minutes prior to its impulsive phase \citep{Veronig2002,Kundu2004,
Joshi2011,Joshi2013,Mitra2020}. While the precursor phase often shows a direct link to the later eruptive phenomenon, the pre-flare activity is viewed as a single or multiple series of small-scale reconnection events within the active region and may indirectly support the eruption by changing the magnetic and plasma conditions favorably \citep{Farnik1996,Farnik1998,Chifor2006,Joshi2011,Joshi2013,Mitra2019}. Arguably the observations of pre-flare or precursor activity have potential to provide insight on the build-up phase of MFR and the triggering mechanism of the subsequent solar eruption. 

During 2015 June 15--29, AR NOAA 12371 passed over the solar visible disk and produced several eruptive flares including geoeffective ones. The long-duration event of GOES class M6.6/H$\alpha$ importance 2B occurred on 2015 June 22 is of particular interest in view of the highly eventful and extended pre-flare phase, its dual peak main phase, as well as the very distinct observations of the MFR structure and overlying strapping field in the AIA EUV filtergrams. This event has been the subject of several studies. \citet{Jing2017} reported on the large-scale dynamics associated with the flare. They noted propagation of footpoint brightening driven by injection of non-thermal particles and the apparent slippage of loops governed by plasma heating and subsequent cooling. \citet{Wang2018} studied the changes in photospheric flows and magnetic field structures associated with the flare.  Their study reveals the role of back reaction of the coronal fields as caused by the flare energy release. \citet{Awasthi2018} analyzed the pre-flare configuration and identified a multiple braided flux rope along the PIL with different degrees of coherency over the pre-flare phase. \citet{Liu2018} analyzed the changes in the photospheric vector magnetic field, which are related to the motion of the flare ribbons. \citet{Kang2019} reported the involvement of ideal instabilities (double arc instability and torus instability) and tether-cutting mechanism as plausible cause of the eruption of the flux rope and subsequent M6.6 flare.   

In this study, we revisit SOL2015-06-22T18:23 to investigate the processes occurring during the extended period prior to the onset of eruption, the pre-flare activity while the quasi-stable MFR continued to build up, and how these processes relate to the subsequent impulsive phase when the MFR underwent spectacular eruption that led to a fast halo CME. The paper is organized as follows. In Section~\ref{sec:observation_data}, we provide a brief discussion about observational data and techniques. Section~\ref{sec:multi-wavelength_results} provides an extensive exploration of the multi-wavelength data in (E)UV, X-ray, and optical bands along with analysis of photosphetic magnetograms. The results are discussed in Section~\ref{sec:discussion}.

\newpage
\section{Observational data and techniques}
\label{sec:observation_data}

This study is primarily based on data from the Atmospheric Imaging Assembly \citep[AIA;][]{Lemen2012} and Heliosesmic and Magnetic Imager \citep[HMI;][]{Schou2012} on board the Solar Dynamics Observatory \citep[SDO;][]{Pesnell2012}. AIA records full-disk images of the corona and transition region up to 0.5 $R_\odot$ above the photosphere in EUV and UV filters. It produces narrow band images centered on specific lines corresponding to seven EUV passbands: They are 94 \AA\ (Fe\begin{scriptsize}XVIII\end{scriptsize}\hspace{-0.13cm}), 131 \AA\ (Fe\begin{scriptsize}VIII\end{scriptsize}\hspace{-0.13cm},\begin{scriptsize}XXI\end{scriptsize}\hspace{-0.13cm}), 171 \AA\ (Fe\begin{scriptsize}IX\end{scriptsize}\hspace{-0.13cm}), 193 \AA\ (Fe\begin{scriptsize}XII\end{scriptsize}\hspace{-0.13cm},\begin{scriptsize}XXIV\end{scriptsize}\hspace{-0.13cm}), 211 \AA\ (Fe\begin{scriptsize}XIV\end{scriptsize}\hspace{-0.13cm}), 304 \AA\ (He\begin{scriptsize}II\end{scriptsize}\hspace{-0.13cm}), and 335 \AA\ (Fe\begin{scriptsize}XVI\end{scriptsize}\hspace{-0.13cm}). UV observations are made at 1600 \AA\ (C\begin{scriptsize}IV\end{scriptsize}\hspace{-0.12cm}) and 1700 \AA\ (nearby continuum). AIA produces 4096 $\times$ 4096 pixel images at a pixel resolution of 0$\farcs$6 pixel$^{-1}$ with a temporal cadence of 12 s for the EUV filters and 24 s for the UV filters. In this study, we have extensively analyzed the observations taken in the 94 \AA\ (log T=6.8) and the 304 \AA\ (log T=4.7) channels besides examining images at other AIA channels. 
 
HMI provides full disk measurements of the intensity, doppler shift, line-of-sight (LOS) magnetic field, and vector magnetic field at the solar photosphere using the 6173~\AA\ Fe I absorption line. Images are produced with 4096 $\times$ 4096 pixel at a pixel resolution of  0$\farcs$5 pixel$^{-1}$ and a temporal cadence of 45 s for velocity, intensity, and LOS magnetic field. For the vector magnetic field, the temporal cadence is 720 s. In order to compare the images from HMI with AIA, we use the SSW routine \textit{hmi\_prep.pro}, which converts the resolution of the HMI images from 0$\farcs$5 pixel$^{-1}$ to 0$\farcs$6 pixel$^{-1}$, which is the pixel resolution of AIA.    

The H$\alpha$ images studied are full disk observations from Big Bear Solar Observatory (BBSO) \citep{Denker1999} with a telescope aperture of 10 cm. These H$\alpha$ observations are taken with a filter of 0.25 \AA\ bandpass centered at the H$\alpha$ line core and 2048$\times$2048 pixel CCD camera. The images have a temporal cadence of $\approx$60 s and pixel resolution $\approx$1$\farcs$0. 

The temporal, spatial, and spectral evolution of the hard X-ray (HXR) emission from the flaring region is analyzed using data from the Reuven Ramaty High Energy Solar Spectroscopic Imager \citep[RHESSI;][]{Lin2002}. RHESSI observed the full Sun with an unprecedented combination of spatial resolution (as fine as $ \sim $2$ {\arcsec} $.3) and energy resolution (1--10~keV) in the energy range 3~keV to 17~MeV. To reconstruct RHESSI HXR images at different energy bands, we have used the CLEAN algorithm \citep{Hurford2002}. For HXR spectroscopy, we generated RHESSI spectra with an energy binning of 1/3 keV from 6 to 15 keV, 1 keV from 15 to 100 keV, and 5 keV from 100 keV onward.  We only used front segments of the detectors, and excluded detectors 2 and 7 (which have lower energy resolution and high threshold energies, respectively; \citet{Smith2002}). The spectra were deconvolved with the full detector response matrix. Two fitting models have been used: line emission from an isothermal plasma and thick-target bremsstrahlung from non-thermal electrons interacting with the chromosphere \citep{Holman2003}. From spectral fits, we derived the temperature (T) and emission measure (EM) of the hot flaring plasma, as well as the non-thermal electron spectral index ($\delta$) for the non-thermal component.

The CME associated with the M6.6 flare under study was observed by the C2 and C3 instruments of the Large Angle and Spectrometric Coronagraph \citep[LASCO;][]{Brueckner1995} on board the Solar and Heliospheric Observatory \citep[SOHO;][]{Domingo1995}. C2 and C3 are white light coronagraphs that image the solar corona with a field-of-view of 1.5 to 6 $R_\odot$ and 3.7 to 30 $R_\odot$, respectively.

To model the coronal magnetic field distribution, we used hmi.sharp$\textunderscore$cea$\textunderscore$720s series vector magnetogram of SDO/HMI as input boundary condition. The magnetogram is remapped using a Lambert cylindrical equal-area projection and presented as (B$_{r}$,B$_{\theta}$,B$_{\phi}$) in heliocentric spherical coordinates corresponding to (B$_{z}$,B$_{y}$,B$_{x}$) in heliographic coordinates \citep{Sun2013}. The magnetogram represents an area of 474 $\times$ 226 pixel$^{2}$ of the AR, which corresponds to an area of 343 $\times$ 163 Mm$^{2}$ on the surface of the Sun. The extrapolation was done upto a height of 163 Mm above the photosphere. To visualize the extrapolated field lines, we used Visualization and Analysis Platform for Ocean, Atmosphere, and Solar Researchers \citep[VAPOR\footnote{\url{https://www.vapor.ucar.edu/}};][]{Clyne2010} software.

\section{multi-wavelength observations and results} 
\label{sec:multi-wavelength_results}

\subsection{Event overview and light curve analysis}   
\label{sec:overview_lightcurve}

\begin{figure}
\epsscale{0.8}
\plotone{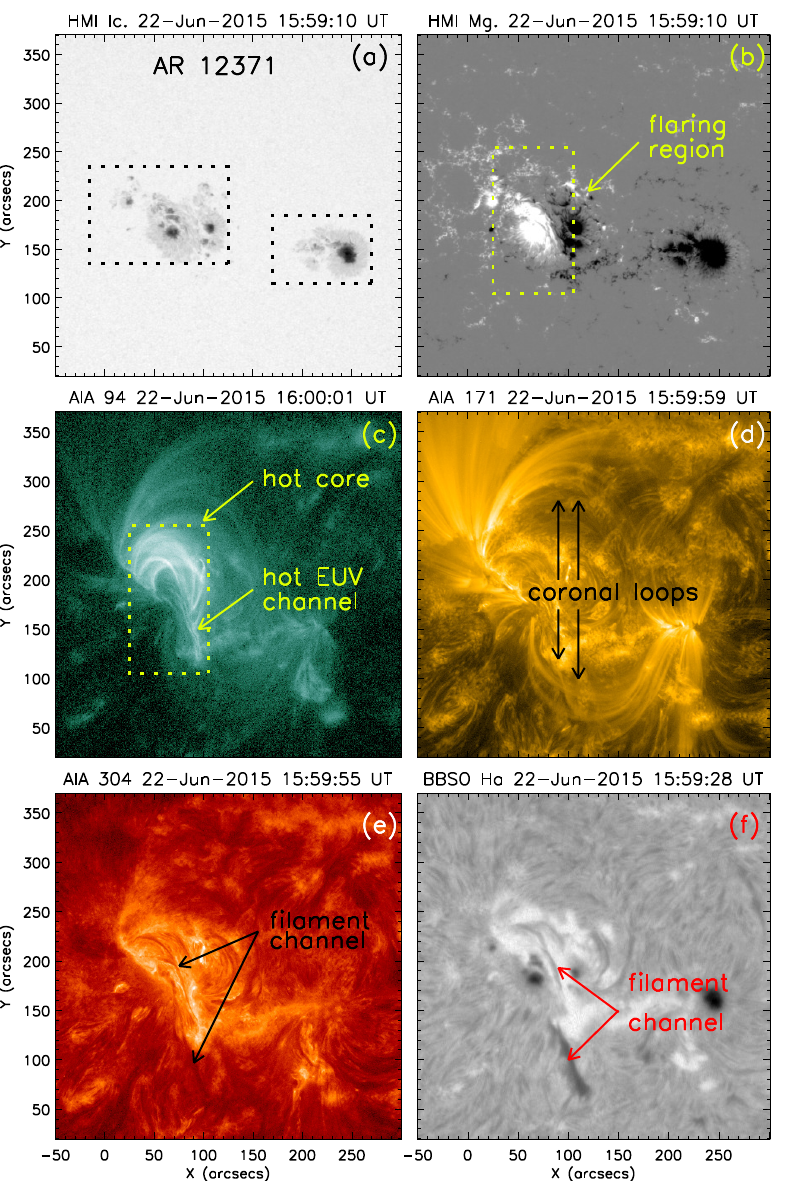}
\caption{Multi-wavelength view of active region NOAA 12371 on 2015 June 22. 
(a) White light image of the active region showing configurations of leading and trailing sunspot groups, which are shown by dotted boxes.
(b) HMI line-of-sight (LOS) magnetogram showing the photospheric magnetic structure of the active region. The flare under investigation primarily originated in the trailing part of the active region. 
(c) AIA 94 \AA\ image of the pre-flare phase showing the hot core region where the M6.6 class flare occurred. (d) AIA 171 \AA\ image showing high coronal loops that lie over the sunspot groups.
(e) AIA 304 \AA\ image showing faint filament structure in the chromospheric level. 
(f) BBSO H$\alpha$ image showing clear filament channel above polarity inversion line (PIL). Comparison of panels (c), (e), and (f) reveals that, a filament exists in the chromosphere underneath the hot EUV channel.  
\label{fig:overview}}
\end{figure}

We investigate an M6.6 class flare from AR NOAA 12371 on 2015 June 22 from 16:00 UT to 23:00 UT. The active region was situated at heliographic coordinate $\approx$N12W08 during the onset of the flare. In Figure \ref{fig:overview}, we present a multi-wavelength view of the active region to compare its morphology at different atmospheric layers of the Sun. The white light image of the active region shows two distinct sunspot groups (shown by dashed boxes in Figure \ref{fig:overview}(a)). A comparison of white light image with LOS magnetogram of the active region suggests that the leading sunspot group is of negative polarity and the trailing sunspot group is comprised of mixed polarity regions making it a $\beta\gamma$ type active region (cf. Figures \ref{fig:overview}(a) and \ref{fig:overview}(b)). The flaring site is located over the trailing sunspot group (shown by yellow dashed box in Figure \ref{fig:overview}(b)). The AIA 94 \AA\ image during the pre-flare phase (Figure \ref{fig:overview}(c)) suggests that, the activity site was associated with intensely emitting closed loops, which we mark by a dotted rectangle and annotate as hot core. Furthermore, we identify a hot channel-like structure at low coronal heights (marked by yellow arrow as hot EUV channel). By examining the images in the AIA 94 \AA\ channel prior to the event over several hours, we find that, the hot channel pre-existed at least $\approx$5.5 hours before the eruptive flare. A comparison of the AIA 94 \AA\ image with a co-temporal HMI magnetogram suggests that the brightest part of the core region with dense coronal loops essentially lie over the trailing part of the active region showing a complex bipolar magnetic distribution of sunspots.
The 171 \AA\ image in Figure \ref{fig:overview}(d) shows high coronal loops connecting the leading and trailing sunspot groups. In Figure \ref{fig:overview}(e), the AIA 304 \AA\ image shows the signature of a filament channel. In Figure \ref{fig:overview}(f), the BBSO H$\alpha$ image clearly shows the filament channel over polarity inversion line (PIL) (cf. Figures \ref{fig:overview}(b) and (f)). We infer the hot EUV channel to be the coronal counterpart of the chromospheric filament delineating the PIL of the trailing bipolar part of the active region.

\begin{figure}    
\epsscale{0.95}
\plotone{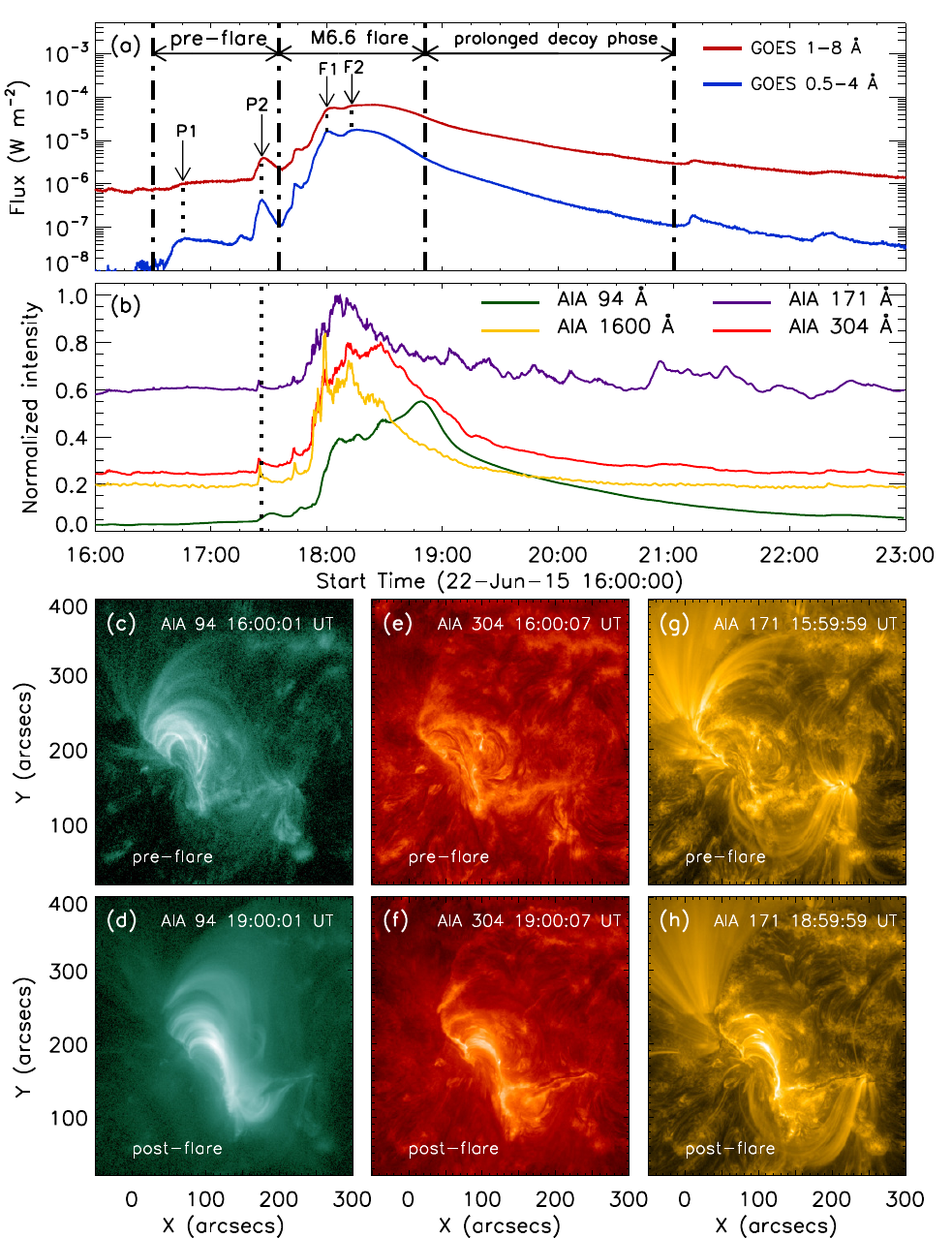}     
\caption{Panel (a): GOES soft X-ray flux in 1--8 \AA\ and 0.5--4 \AA\ channel from 16:00 UT to 23:00 UT on 2015 June 22. We find two stages in the pre-flare phase that peak at 16:45 UT (marked as P1) and 17:26 UT (marked as P2), respectively. We also observe dual flare-peak structure in the main phase of the M6.6 flare, indicated as F1 and F2 at 18:00 UT and 18:13 UT, respectively. Panel (b): AIA light curves normalized by peak intensity of respective AIA filters. For clear view, light curves have been scaled by factors of 0.55 and 0.8 for 94 \AA\ and 304 \AA\ channels, respectively. The peak P2 in GOES soft X-ray light curves in the pre-flare phase corresponds to a peak in AIA light curves, which is shown by dotted line. We readily observe that, the structure of the active region shows significant changes during the course of the flare. As a comparison between pre- and post-flare phases, we plot the active region corona in AIA 94 \AA\ (cf. panels (c) and (d)), 304 \AA\ (cf. panels (e) and (f)), and in 171 \AA\ (cf. panels (g) and (h)).\\
(An animation of this figure showing the temporal evolution of the flare is available in the online material.)}

\label{fig:lightcurve}
\end{figure}

\begin{deluxetable*}{cccp{7cm}}
\tablenum{1}
\tablecaption{Summary of different phases of M6.6 flare\label{tab:summary}}
\tablewidth{1pt}
\tablehead{
\colhead{Serial} & \colhead{Phases} & \colhead{Duration (UT)} &
\colhead{Remarks}  \\
\colhead{No.} & \colhead{} & \colhead{} &
\colhead{} 
}
\startdata
1 & Pre-flare phase & 16:30 -- 17:35  & Two distinct peaks (P1 and P2) are observed 
in GOES SXR light curves at $\approx$16:45 UT and $\approx$17:26 UT.\\
2 & M6.6 flare &  17:35 -- 18:51  & A distinct subpeak at $\approx$17:44 UT in X-ray light curves (GOES and RHESSI) during the rise phase (17:35 -- 18:00 UT); broad maximum phase with dual peak structures (F1 and F2) in GOES light curves at $\approx$18:00 UT and $\approx$18:13 UT; eruption of hot channel begins at $\approx$17:40 UT; CME first detected in LASCO C2 coronagraph at $\approx$18:36 UT.\\
3 & Post-flare phase & 18:51 -- 21:00 & Very gradual decline of SXR emission in GOES light curves for $\approx$2 hours; after which the SXR flux reached to pre-flare background level; emission from large post-flare loops.\\
\enddata
\end{deluxetable*}

\begin{figure}
\includegraphics[width=\textwidth,scale=2.2]{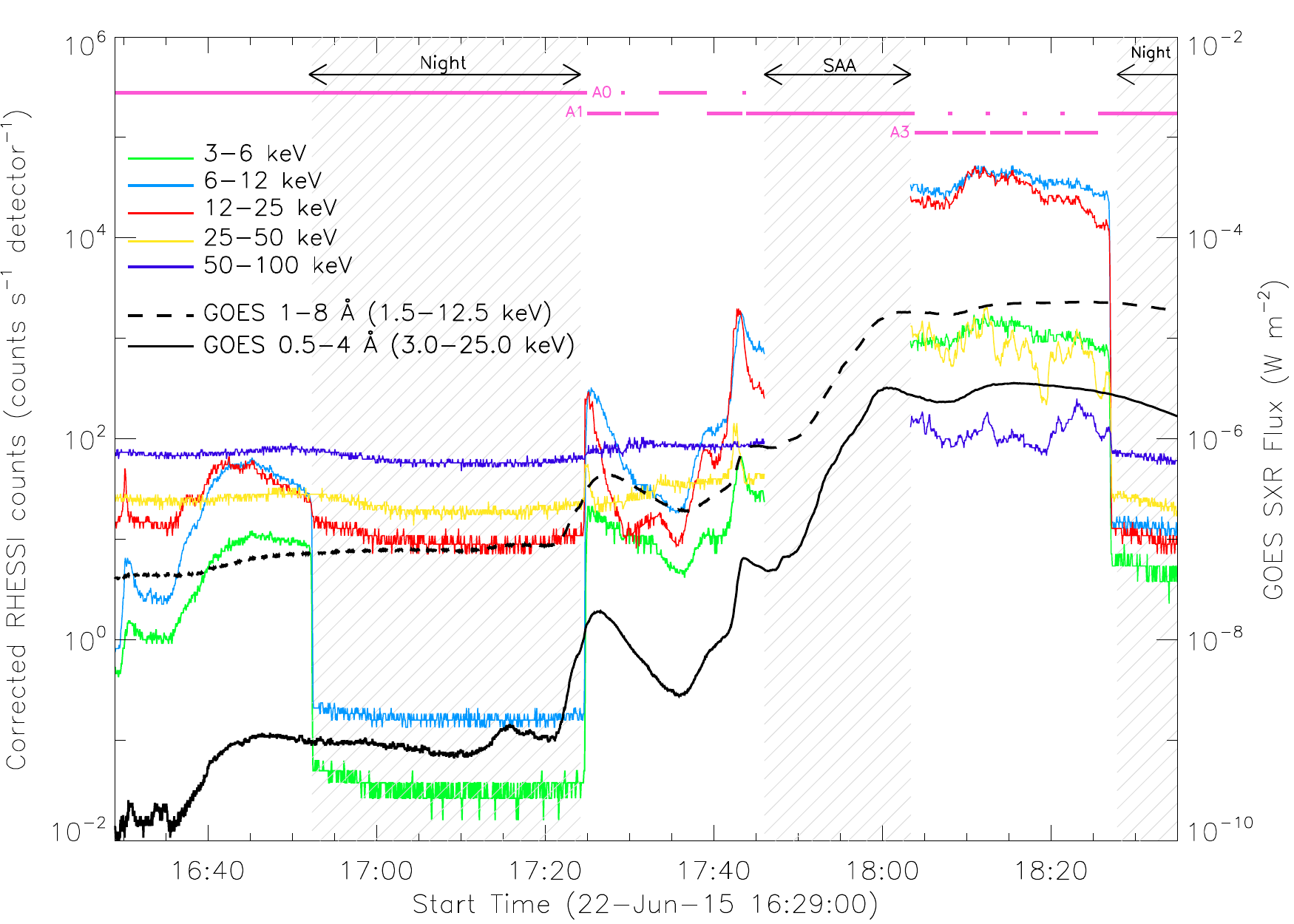}
\caption{Temporal evolution of X-ray count rates observed by RHESSI from 16:29 UT to 18:35 UT in energy bands of 3--6, 6--12, 12--25, and 25--50 keV with a time cadence of 4 s. GOES SXR light curves in 1--8 \AA\ and 0.5--4 \AA\ channels are also shown by dashed and solid lines, respectively. The hatched regions denote unavailability of solar X-ray data due to RHESSI night (N) and South Atlantic Anomaly (SAA). Different attenuator states (A0, A1, and A3) are shown by horizontal bars at the top.}
\label{fig:rhessi_lc}
\end{figure}

\begin{figure}[h]
\includegraphics[width=\textwidth,scale=2.2]
{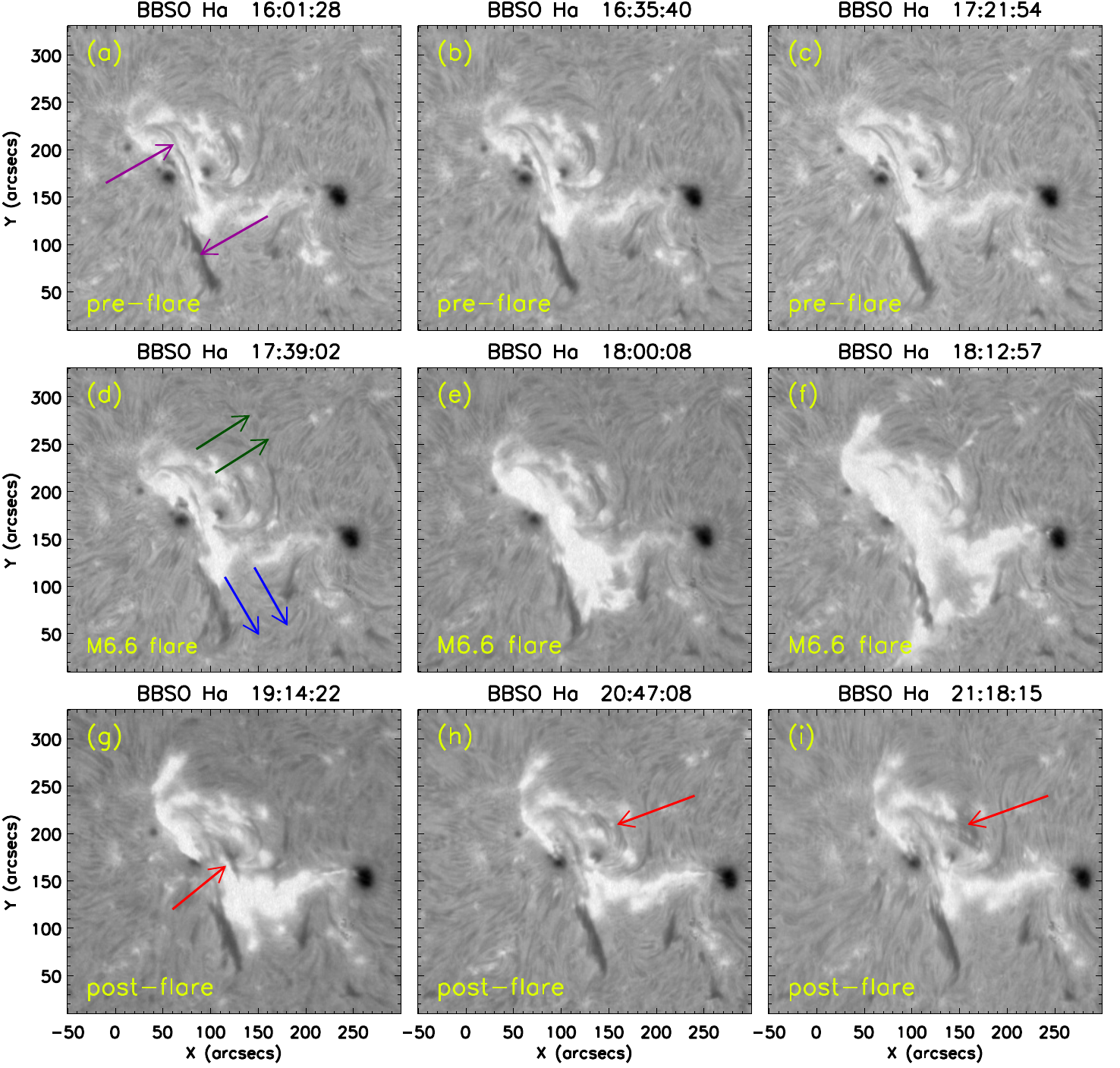}
\caption{BBSO H$\alpha$ filtergrams showing the temporal evolution of different phases of M6.6 flare, namely, pre-flare phase, main phase, and post-flare phase in panels (a)--(c), (d)--(f), and (g)--(i), respectively. Two distinct parts of the filament (shown by purple arrows in panel (a)), together constitute a filament channel. The onset of M6.6 flare is preceded by activation of the filament channel in two different directions (shown by blue and green arrows in panel (d); For details, see Section \ref{sec:overview_lightcurve}). An upward motion of filament material is observed during the post-flare phase, which is shown by red arrows in panels (g)--(i).\\
\label{fig:halpha}}
\end{figure}

\begin{figure}
\includegraphics[width=\textwidth,scale=0.8]{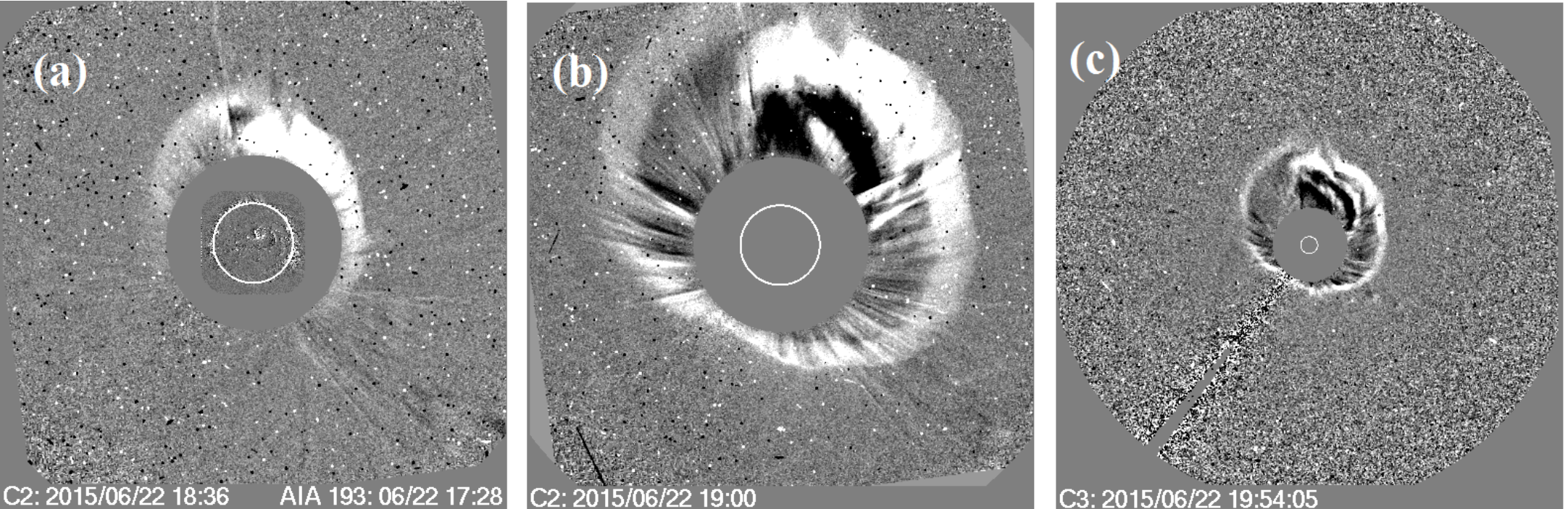}
\caption {Running difference images of LASCO C2 (panels (a) and (b)) and C3 (panel (c)) coronagraph. Panel (a) shows first detection of CME in C2 coronagraph. A full disk image of the Sun in AIA 193 \AA\ is overplotted on the coronagraph occulter. The CME was first detected in C3 coronagraph at $\approx$18:54 UT\hspace{0.05cm}\textcolor{blue}{$^{3}$}.}

$^{3}$\hspace{0.05cm}\url{https://cdaw.gsfc.nasa.gov/movie/make_javamovie.php?stime=20150622_1708&etime=20150622_2105&img1=lasc2rdf&title=20150622.183605.p358g;V=1209km/s}

\label{fig:cme}
\end{figure}

The GOES soft X-ray (SXR) light curves in 1--8 \AA\ and 0.5--4 \AA\ (Figure \ref{fig:lightcurve}(a)), show distinct pre-flare, main, and gradual phases of the long duration flare event under study. We see two distinct peaks in the pre-flare phase (P1 and P2) at $\approx$16:45 UT and $\approx$17:26 UT. A sharp rise in the GOES SXR flux at $\approx$17:35 UT indicates the start of main phase of the M6.6 flare with dual peak structures (F1 and F2) at $\approx$18:00 UT and $\approx$18:13 UT. According to the GOES flare catalog, which is based on the return of the SXR flux to half of its peak value, the flare lasted till 18:51 UT. However, the GOES profiles (Figure \ref{fig:lightcurve}(a)) clearly reveal enhanced soft X-ray emission from the flaring region for several hours ($\approx$up to 21:00 UT), that we mark as prolonged decay phase. In Table \ref{tab:summary}, we summarize the different phases of the flare evolution along with their characteristics, which we discuss in subsequent sections.   

Normalized intensity light curves of the AIA channels 171~\AA, 304~\AA, 94~\AA, and 1600~\AA\ filters are shown in Figure \ref{fig:lightcurve}(b). In general, these (E)UV light curves show similar trends than the GOES SXR light curves with some time delays in the peak emission among different bands. The first peak P1 of the GOES light curve in 0.5--4 \AA\ channel is not seen in the AIA light curves; while the second peak P2 is clearly visible (shown by dotted lines in Figure \ref{fig:lightcurve}(b)). The 94 \AA\ light curve shows a significant time-shift in the peak compared to other AIA light curves and continued emission for a longer period (up to 21:00 UT). The AIA 171 \AA\ light curve shows significant variability in the decay phase, which is not seen in other AIA light curves. The 1600 \AA\ light curve shows dual peak structure in the main phase of the M6.6 flare similar to GOES light curves.

In Figure 2, we provide a comparison of the active region corona during the pre- and post-flare phases of the flare as recorded in different SDO/AIA channels; namely: in 94 \AA\ (cf. Figures 2(c)--(d)), in 304 \AA\ (cf. Figures 2(e)--(f)), and in 171 \AA\ (cf. Figures 2(g)-(h)). In the pre-flare stage, we observe low lying coronal loops with a faint signature of a hot coronal channel underlying the low coronal loops in the 94 \AA\ image (Figure \ref{fig:lightcurve}(c)). In the post-flare stage, dense and bright post-flare loops are observed to be formed (Figure \ref{fig:lightcurve}(d)). In AIA 304 \AA, we observe the signature of filament in pre-flare stage (Figure \ref{fig:lightcurve}(e)), whose eruption gives rise to formation of post-flare loop arcades (Figure \ref{fig:lightcurve}(f)). In the post-flare stage, AIA 171 \AA\ observations also reveal the formation of dense post-flare loop arcades (Figure \ref{fig:lightcurve}(h)).       

In Figure \ref{fig:rhessi_lc}, we show RHESSI X-ray count rates in different energy bands from 3--100 keV in the interval of 16:30 UT to 18:35 UT. The co-temporal GOES SXR light curves in 1--8 \AA\ and 0.5--4 \AA\ are overplotted on the RHESSI count rates. We observe simultaneous occurrence of peaks in GOES SXR and lower energy RHESSI ($<$25 keV) light curves in the pre-flare stage (from $\approx$16:30 UT to $\approx$17:35 UT). In the main phase of the flare (from $\approx$17:35 UT onward), the higher energy RHESSI ($>$25 keV) light curves show distinct small peaks.

To show the overall evolution of the M6.6 flare, we present a few representative H$\alpha$ images in Figure \ref{fig:halpha}, which are obtained from BBSO. At the pre-flare stage, we observe the presence of a filament channel along the PIL of the active region (shown by purple arrows in Figure \ref{fig:halpha}(a)). During the early stage of the main phase of the flare, we observe activation of the filament channel in two different directions: along the length of the channel toward southwest (shown by blue arrows in Figure \ref{fig:halpha}(d)) and along northwest direction (shown by green arrows in Figure \ref{fig:halpha}(d)). The filament continued to erupt along the northwest direction and produced intense flare brightening (Figures \ref{fig:halpha}(e)--(f)). The two-step activation of the filament channel may be associated with the pre-flare activities that led to the partial eruption of the filament. In the post-flare phase, we identify a filamentary material to be erupted from the core part of the flaring region, which is shown by red arrows in Figures \ref{fig:halpha}(g)--(i).

The eruption of the hot channel and the subsequent flare resulted into a fast halo CME, which was observed by C2 and C3 coronagraphs of LASCO on-board SOHO (Figure~\ref{fig:cme}). Various CME parameters have been gathered from LASCO CME catalog \citep{Yashiro2004}\footnote{\url{https://cdaw.gsfc.nasa.gov/CME_list/UNIVERSAL/2015_06/univ2015_06.html}}. The CME was first detected by C2 at  $\approx$18:36 UT (Figure \ref{fig:cme}(a)) at the height of 4.1 $R_{\odot}$ and propagated with a projected linear speed of $\approx$1200 km s$^{-1}$ measured at position angle 357$^{\circ}$.

\subsection{Structure and evolution of the photospheric magnetic field}
\label{sec:AR_magnetic_field}
\begin{figure}
  
\epsscale{0.95}
\plotone{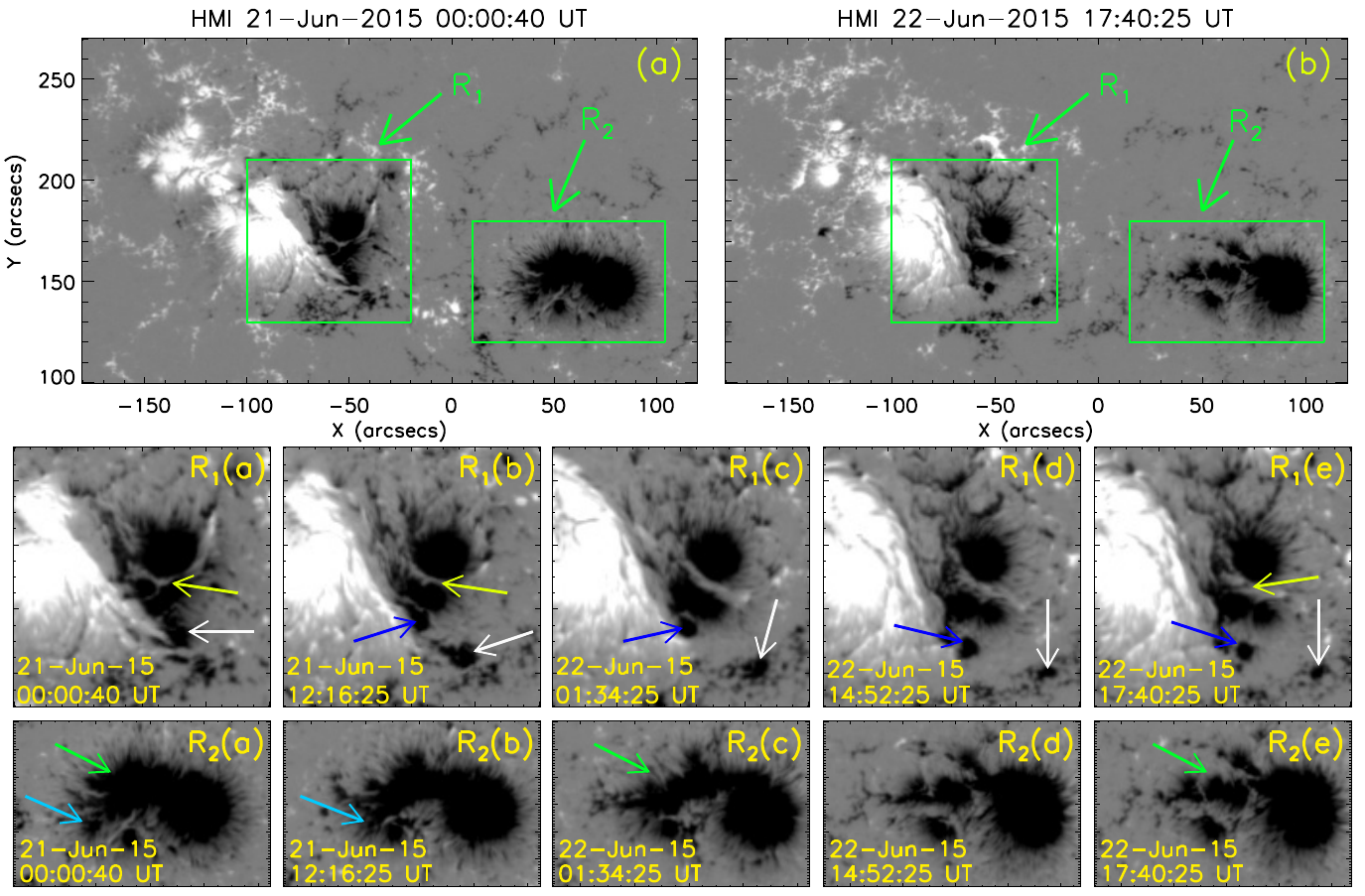} [h!]
\caption{Panels (a)--(b): HMI LOS magnetograms of the active region NOAA 12371 at 00:00~UT on 2015 June 21 (i.e., one day before the studied event) and 17:40 UT on 2015 June 22 (i.e., in the beginning of the M6.6 flare) are plotted to show the changes in the photospheric magnetic structures. We mark two sub-regions R$_{1}$ and R$_{2}$ in magnetograms that exhibited significant changes. In panels R$_{1}$(a)--R$_{1}$(e), we show a few representative snapshots of the sub-region R$_{1}$ to highlight important changes. Yellow arrows indicate a light bridge dividing the negative polarity region of the trailing sunspot group, which apparently underwent rotation in clockwise direction and eventually became thicker. White arrows indicate motion of a small negative region toward southwest. Blue arrows indicate southward motion of another negative polarity region. In panels R$_{2}$(a)--R$_{2}$(e), we show the evolution of the sub-region R$_{2}$. The sky blue arrow in panel R$_2$(a) shows a small negative polarity region, which merged into the bigger negative polarity region (cf. panels R$_2$(a) and R$_2$(b)). Green arrows in panels R$_2$(a), R$_2$(c), and R$_2$(e) indicate a region where the magnetic flux rapidly evolved and eventually resulted in significant cancellation of negative flux.\
\\(An animation of this figure is available in the online material.)}
    
\label{fig:magnetic_AR_evolution}
\end{figure}

Coronal magnetic configurations are deeply associated with changes of photospheric magnetic structures. Therefore, to understand the cause of flaring activities and filament eruptions, it is essential to study the changes associated with the photospheric magnetic fields. The magnetic structure of the AR 12371, one day prior to the event under investigation (i.e., 21 June 2015), is shown in Figure~\ref{fig:magnetic_AR_evolution}(a). The magnetic structure of the same AR just before the event is shown in Figure~\ref{fig:magnetic_AR_evolution}(b). A comparison of panels (a) and (b) reveals that the two sub-regions R$_1$ and R$_2$ underwent significant changes (see animation attached with Figure \ref{fig:magnetic_AR_evolution}). In order to further probe the magnetic changes, a few representative magnetograms of the sub-region R$_1$ from 2015 June 21 to 2015 June 22 are shown in panels R$_1$(a)--R$_1$(e) of Figure \ref{fig:magnetic_AR_evolution}. A light bridge (marked by yellow arrow) is noticeable, which gradually underwent apparent rotation in clockwise direction. This light bridge separated the negative polarity sunspot and showed an increase in its width. White arrows show motion of a negative polarity region toward southwest (cf. Figures \ref{fig:magnetic_AR_evolution}R$_1$(a)--R$_1$(e)). Blue arrows show southward motion of another negative polarity region within sub-region R$_1$.    
 
In the first two panels of the LOS magnetograms of the R$_2$ region, the sky blue arrows show an island of negative polarity region, which eventually merged into the major negative polarity region north of it (cf. Figures~\ref{fig:magnetic_AR_evolution}R$_2$(a)-R$_2$(b)). Notably, the eastern part of R$_{2}$ showed a very intriguing dynamical evolution with multiple events of fragmentation and merging of magnetic structures. Eventually the region exhibited significant cancellation of negative magnetic flux (cf. region marked by green arrows in Figures~\ref{fig:magnetic_AR_evolution}R$_2$(a), R$_2$(c), and R$_2$(e)).

\begin{figure}
\epsscale{0.75}
\plotone{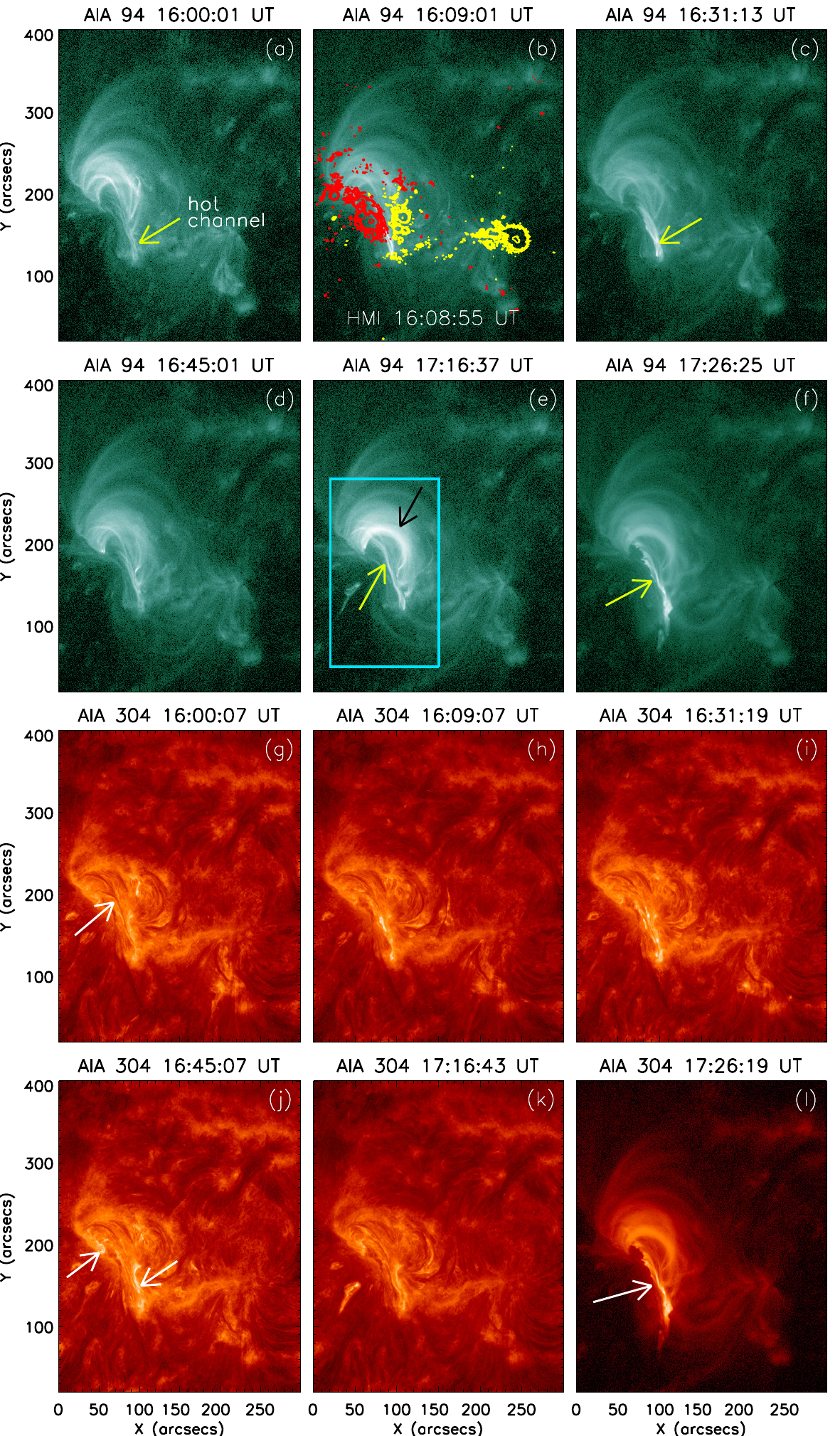}  
\caption{Pre-flare phase of M6.6 flare shown in AIA 94 \AA\ and 304 \AA\ image sequences. Panels (a)--(f): Sequence of AIA 94 \AA\ images showing activation and pre-eruption stages of the hot channel (marked by yellow arrows in panels (a), (c), (e), and (f)) and overlying coronal loops (marked by black arrow in panel (e)). Panel (b) shows overplotted co-temporal HMI LOS magnetogram. The positive and  negative polarities are shown by red and yellow contours respectively, with contours levels set as $\pm$[500, 800, 1000, 2000] G. The box in panel (e) indicates the field-of-view of the images plotted in Figure \ref{fig:RHESSI_AIA94_pre-flare}. Panels (g)--(l): Simultaneous imaging in the AIA 304 \AA\ channel. A filament structure is shown by white arrow in panel (g). White arrows in panel (j) show the appearance of two brightenings on the two sides of the filament channel. The white arrow in panel (l) shows enhanced brightening from the filament channel.}

\label{fig:pre-flare}
\end{figure}

\begin{figure}
\epsscale{0.98}
\plotone{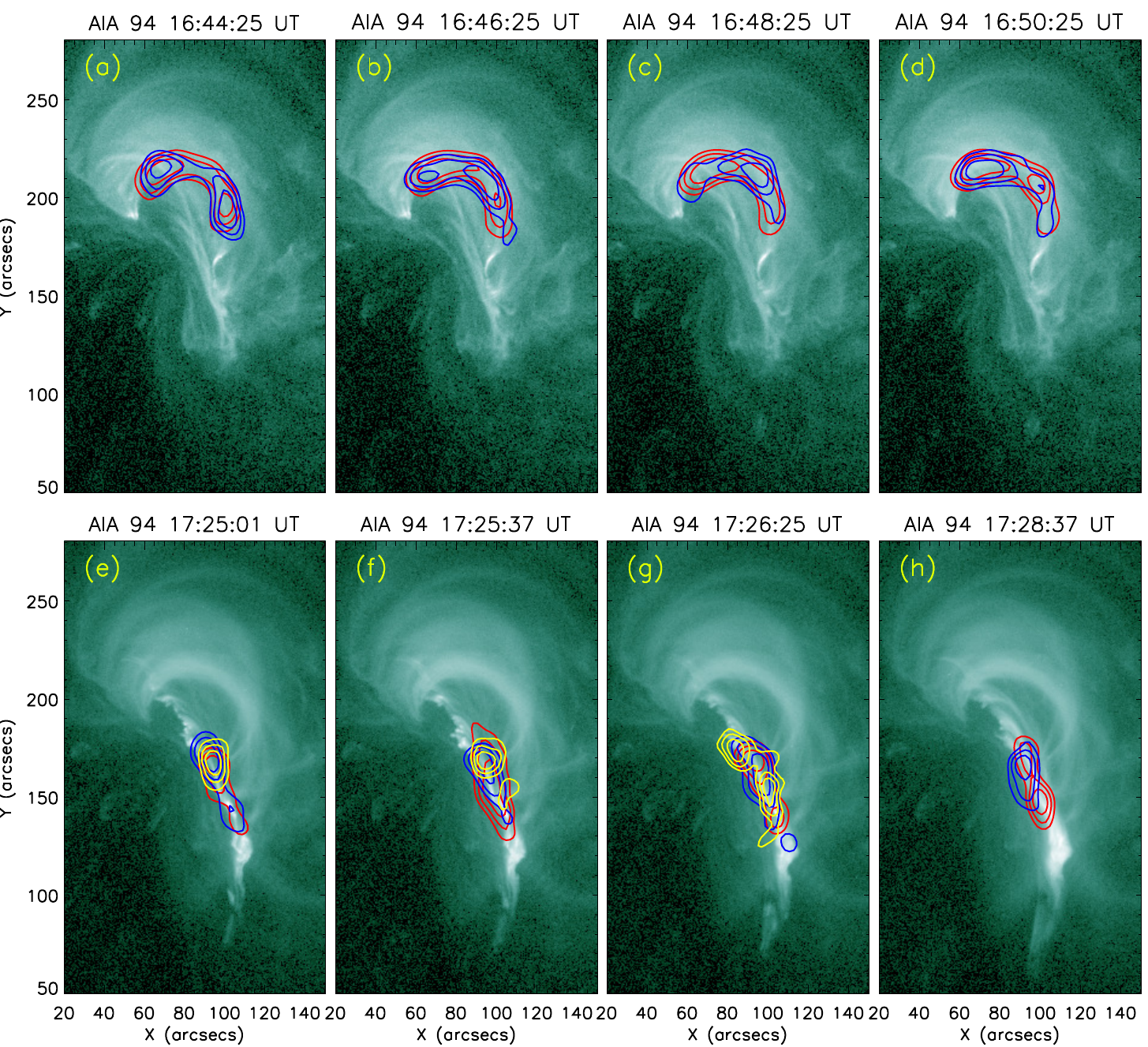}
\caption{Sequence of RHESSI X-ray images in 5--10 keV (red contours), 10--15 keV (blue contours), and 15--25 keV (yellow contours) overplotted on co-temporal AIA 94 \AA\ images. Panels (a)--(d):  sequence of images for first stage (peaked at P1) of the pre-flare phase,  where the X-ray sources are observed to be emitted from the overlying coronal loops. Panels (e)--(h): sequence of images for second stage (peaked at P2) of the pre-flare phase. In this period X-ray emissions are observed from the low-lying hot EUV channel below the coronal loops. The X-ray images are reconstructed by the CLEAN algorithm with integration time of 40 seconds. The contours drawn are at 70\%, 80\%, and 90\% of the peak flux in each image.}
\label{fig:RHESSI_AIA94_pre-flare}
\end{figure}

\begin{figure}
  
\epsscale{0.95}
\plotone{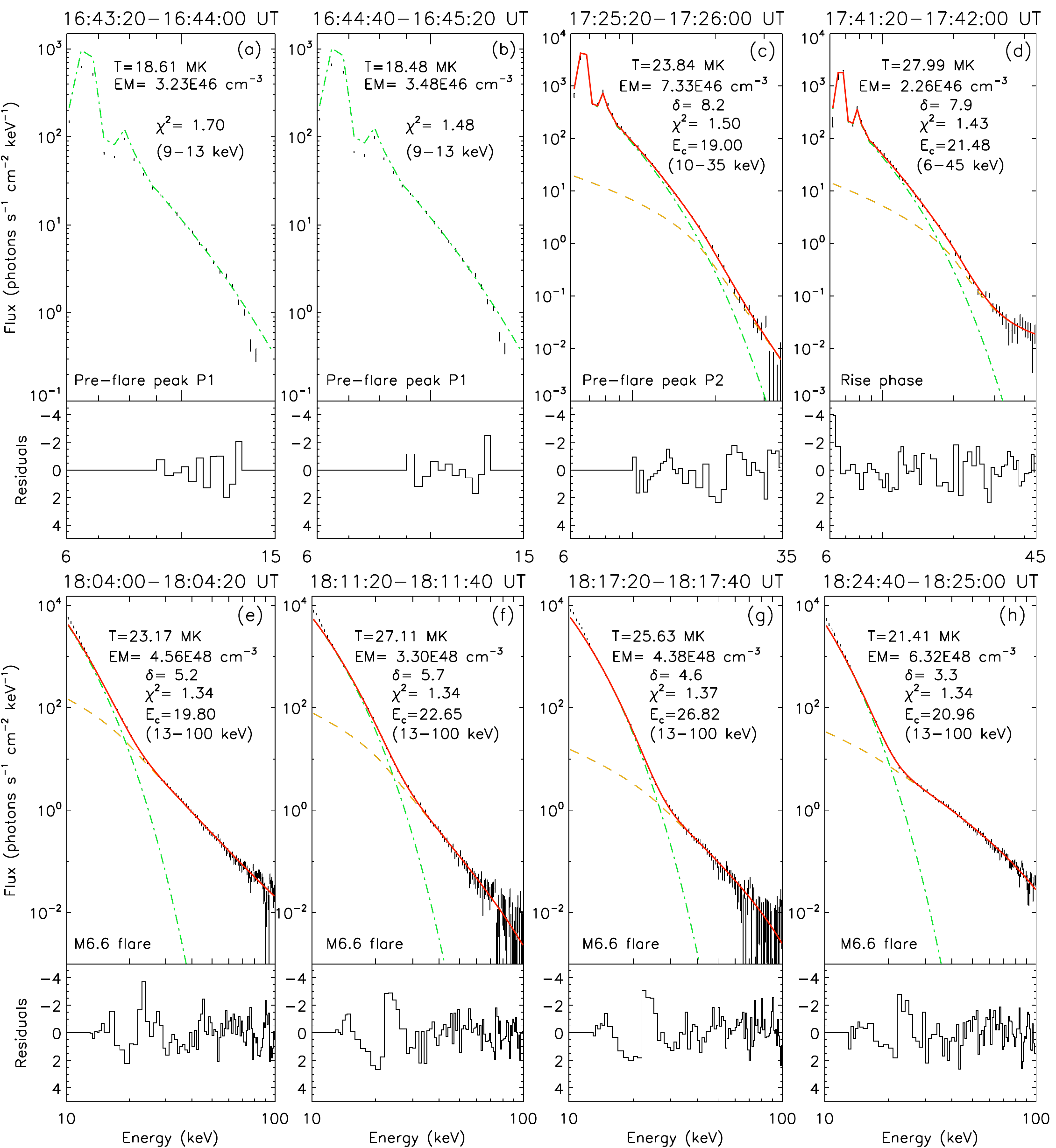}
\caption{X-ray spectral fit of RHESSI data during various phases of the M6.6 flare. Panels (a) and (b) show spectral fit during the peak P1 ($\approx$16:45 UT) and panel (c) shows spectral fit during the peak P2 ($\approx$17:26 UT). We note that, thermal emission is dominant during the peak P1, whereas during the peak P2, we observe appearance of non-thermal component in the spectral fit. Both temperature and emission measure rises during the peak P2 compared to P1. Temporal evolution of spectral fit parameters in the main phase of the M6.6 flare is shown in panels (e)--(h).}
\label{fig:RHESSI_spectra}
\end{figure}

\begin{figure}
\epsscale{0.95}
\plotone{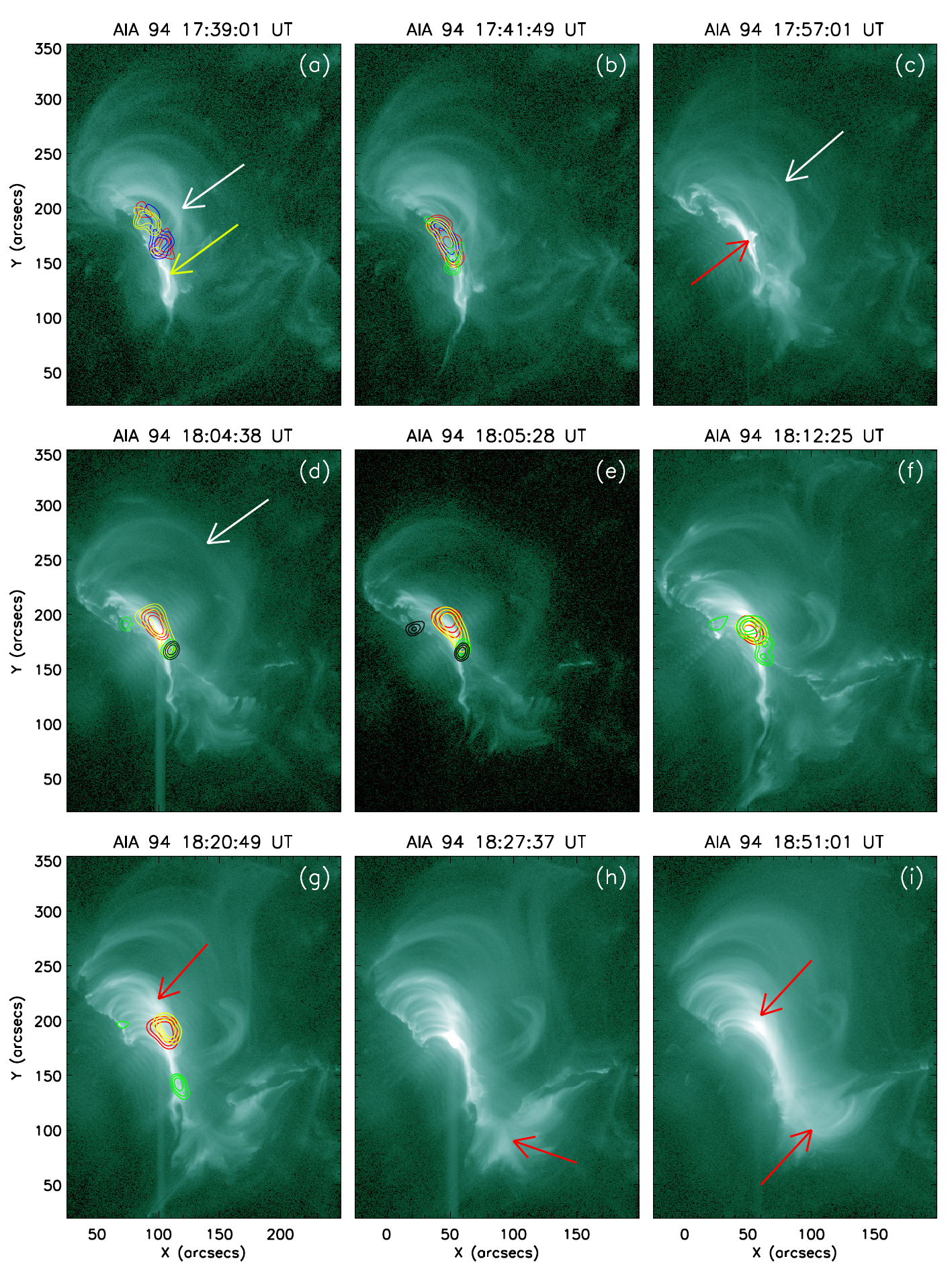}
           
\caption{Sequence of AIA 94 \AA\ images showing evolutionary phases of the eruption of hot channel and associated M6.6 flare. Panel (a) shows hot EUV channel (marked by yellow arrow) and overlying coronal loops (marked by white arrow). The erupting front of the hot channel is shown by white arrows in panels (c) and (d). Red arrow in panel (c) shows start of formation of post-flare loops. RHESSI images in 5--10 keV (red contours), 10--15 keV (blue contours), 15--25 keV (yellow contours), 25--50 keV (green contours), and 50--100 keV (black contours) are reconstructed by CLEAN algorithm with integration time of 32 seconds. The contour levels are set as 70\%, 80\%, and 90\% of the peak flux in each image. Panels (g)--(i) show formation of post-flare loop arcades. The red arrow in panel (g) shows the post-flare loops in the northern part of the flaring region, which ultimately converts into dense post-flare loop arcades. Red arrow in panel (h) shows the start of formation of post-flare loops in the southern part of the flaring region. In panel (i), dense post-flare loop arcades in both northern and southern part of the flaring region are indicated by red arrows.}

\label{fig:RHESSI_AIA94_impulsive}
\end{figure}

\subsection{Build-up and activation of the hot coronal channel}
\label{sec:MFR_activation}

Based on the GOES light curves (Figure \ref{fig:lightcurve}(a)), we have defined the pre-flare phase from $\approx$16:30 UT to $\approx$17:35 UT (see Table \ref{tab:summary}). In Figure \ref{fig:pre-flare}, we show EUV images of the active region during the pre-flare phase by a few representative AIA 94~\AA\ and 304~\AA\ images. Initially, a faint hot channel is identified beneath coronal loops in the 94~\AA\ images (marked by yellow arrow in Figure \ref{fig:pre-flare}(a)). From the HMI LOS magnetogram contours overplotted on the 94 \AA\ image (Figure \ref{fig:pre-flare}(b)), it becomes clear that the hot channel lies over the PIL formed within the trailing sunspot group. Subsequently, the brightening of the hot channel intensifies (marked by yellow arrows in Figures~\ref{fig:pre-flare}(c) and \ref{fig:pre-flare}(e)) and it appears distinctly different from the surrounding regions. In view of the spatial association of the hot channel and the overlying low coronal loops showing bright emission during the pre-flare phase, we identify the region shown inside the box in Figure \ref{fig:pre-flare}(e) as the active region core and focus on its evolution in Figure \ref{fig:RHESSI_AIA94_pre-flare}. The hot channel continues to show enhanced emission till the occurrence of pre-flare peak P2 in GOES light curves (at $\approx$17:26 UT), which is indicated by yellow arrow in Figure \ref{fig:pre-flare}(f).
 
In AIA 304 \AA\ images, we observe a filament as the chromospheric counterpart of the hot channel (marked by white arrow in Figure \ref{fig:pre-flare}(g)). Subsequently, we observe two parallel ribbon-like brightenings at $\approx$16:45~UT (shown by white arrows in Figure \ref{fig:pre-flare}(j)), which are co-temporal with the pre-flare peak P1 (Figure~\ref{fig:lightcurve}(a)). Thereafter, the flux rope undergoes enhanced brightening at $\approx$17:26 UT (Figure \ref{fig:pre-flare}(l)). We note this brightening to be simultaneous with the appearance of pre-flare peak P2 in GOES light curves (Figure \ref{fig:lightcurve}(a)). 

In Figure~\ref{fig:RHESSI_AIA94_pre-flare}, we show a sequence of AIA 94~\AA\ images with co-temporal RHESSI X-ray images (as contours) overplotted on each panel. The evolution of the active region core during the first peak (P1) of pre-flare phase is shown in Figures \ref{fig:RHESSI_AIA94_pre-flare}(a)--\ref{fig:RHESSI_AIA94_pre-flare}(d). During this phase, the X-ray emissions up to 15 keV is observed to come from the region of overlying coronal loops. We note, the X-ray emissions have spatially extended structure with multiple centroids, being morphologically directly resembling with the coronal loop system of the core region.

During the second peak (P2) of the pre-flare phase, we observe strong X-ray emission from the hot channel (Figures \ref{fig:RHESSI_AIA94_pre-flare}(e)--(h)) with X-ray emissions up to 25 keV. Evolution of the X-ray sources during this period is very striking. Initially at $\approx$17:25 UT, we find X-ray emitting sources with distinct centroids in the energy bands up to 15 keV (Figure \ref{fig:RHESSI_AIA94_pre-flare}(e)). The X-ray emissions in 5--10 keV energy band show nearly double centroid structure throughout the pre-flare peak P2 (Figures \ref{fig:RHESSI_AIA94_pre-flare}(e)--(h)), whereas, the X-ray centroids in 10--15 keV energy band dissolve (Figure \ref{fig:RHESSI_AIA94_pre-flare}(e)--(h)). Notably, the X-ray emitting sources in 15--25 keV energy band show appearance of multiple centroids (Figure \ref{fig:RHESSI_AIA94_pre-flare}(g)), which disappear afterwards (Figure \ref{fig:RHESSI_AIA94_pre-flare}(h)).

\begin{figure}
\epsscale{0.95}
\plotone{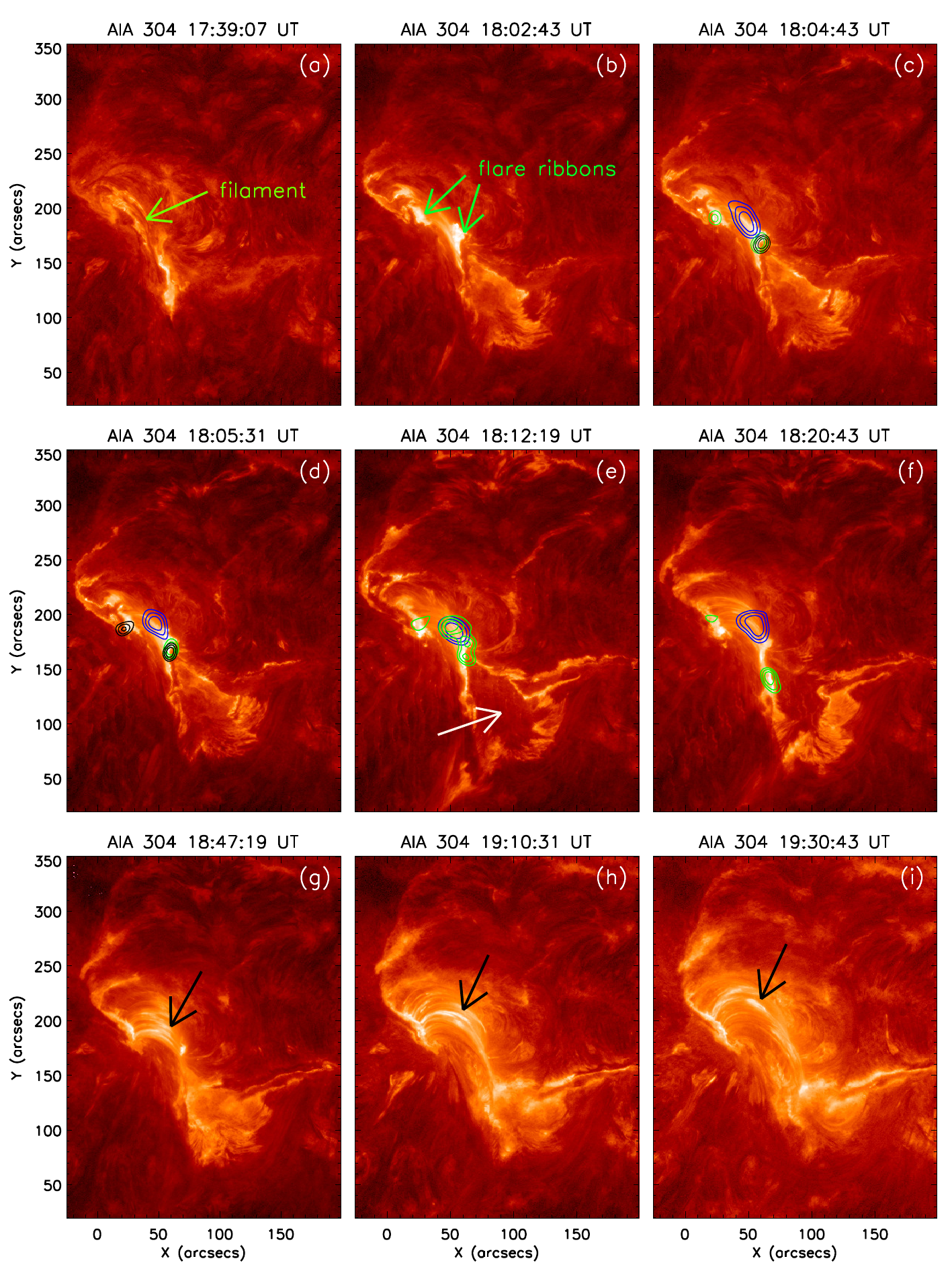}
              
\caption{Sequence of AIA 304 \AA\ images showing eruption of flux rope (i.e., filament) and formation of post-flare loop arcades. The filament is shown by green arrow in panel (a). Subsequently we observe parallel flare ribbons at the footpoints of the erupting filament at $\approx$18:02 UT (marked by green arrows in panel (b)). RHESSI images in 10--15 keV (blue contours), 25--50 keV (green contours), and 50--100 keV (black contours) are reconstructed by CLEAN algorithm with integration time of 32 seconds. The contours denote 70\%, 80\%, and 90\% of peak flux in each image. White arrow in panel (e) marks a region, which is gradually filled by chromospheric brightening (cf. panels (e)--(i)). We observe diffuse emission from post-flare coronal loops (shown by black arrows in panels (g)--(i)).}
\label{fig:RHESSI_AIA304_impulsive}
\end{figure}

In Figure~\ref{fig:RHESSI_spectra}, we present spatially integrated, background subtracted RHESSI spectra along with their respective fits and residuals for a few selected intervals. Panels (a)--(c) of Figure \ref{fig:RHESSI_spectra} correspond to the spectra of pre-flare phase intervals. The RHESSI X-ray spectra during the first pre-flare phase (peaked at P1, see Figure \ref{fig:lightcurve}(a)) shows thermal emission only (Figures \ref{fig:RHESSI_spectra}(a)--(b)). To estimate the characteristics of hot flaring plasma, namely temperature (T) and emission measure (EM), the best spectral fit results are obtained with fitting in the energy range of 9--13 keV for this interval. At this stage, the plasma temperature is $\approx$19 MK and emission measure is $\approx$3$\times$10$^{46}$ cm$^{-3}$. In the second pre-flare phase, which peaks at P2 (Figure \ref{fig:lightcurve}(a)), we find rise in temperature ($\approx$24 MK) as well as emission measure ($\approx$7$\times$10$^{46}$ cm$^{-3}$; Figure \ref{fig:RHESSI_spectra}(c)), which suggests increase of thermal emission along with volume of heated plasma. Notably, during this second phase of the pre-flare activity, the X-ray emission rises to $\approx$35 keV above the background level. Contrary to the first pre-flare phase, the second pre-flare phase shows distinct yet moderate non-thermal emission above 19 keV with a steep electron spectral index ($\delta$) of $\approx$8.2. The spectral fit results obtained during the rise and main phase of the flare are presented in Figures \ref{fig:RHESSI_spectra}(d)--(f), which are discussed at the end of Section \ref{sec:MFR_eruption}.

\begin{figure}

\epsscale{0.55}
\plotone{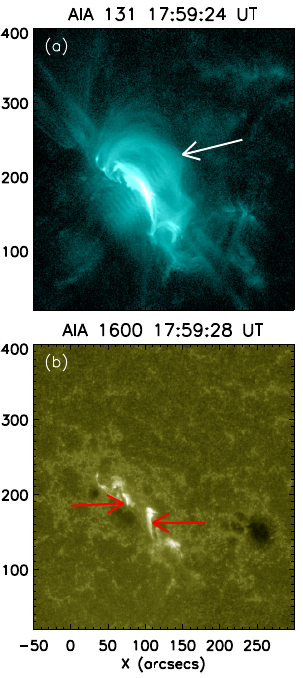}

\caption{Structure of the solar corona and associated active region in EUV (AIA 131 \AA) and UV (AIA 1600 \AA) channels during the peak of the flare. In panel (a), we indicate the erupting hot flux rope structure in 131 \AA\ image by white arrow, while the co-temporal observation in AIA 1600 \AA\ shows conjugate and sheared flare ribbon brightenings, which are shown by red arrows in panel (b). 
\label{fig:AIA_131_1600}}
\end{figure}

\newpage

\subsection{Hot channel eruption and further consequences} 
\label{sec:MFR_eruption}

The main phase of the M6.6 flare is illustrated by a few representative AIA 94~\AA\ images in Figure \ref{fig:RHESSI_AIA94_impulsive}. The activated hot channel (indicated by the yellow arrow in Figure \ref{fig:RHESSI_AIA94_impulsive}(a)) starts to erupt upward, distending the overlying coronal loop system (indicated by white arrow in Figure \ref{fig:RHESSI_AIA94_impulsive}(a)). At this early stage, the X-ray emission is originated at energies $\lessapprox$25 keV and the sources lie in a relatively compact region (Figure~\ref{fig:RHESSI_AIA94_impulsive}(a)). 
With the further upward expansion of the hot channel, we observe emission from conjugate HXR sources of 25--50 keV energies (green contours), which appears to be located near the anchored footpoints of the hot channel (Figure~\ref{fig:RHESSI_AIA94_impulsive}(b)). Importantly, the continuous rise of the GOES flux is superimposed with a distinct peak at $\approx$17:44~UT, which clearly appears in all the high energy RHESSI X-ray light curves up to 50 keV energies (Figure~\ref{fig:rhessi_lc}). The hot channel rises gradually with clear and intact structure visual in direct AIA 94~\AA\ images. In Figures~\ref{fig:RHESSI_AIA94_impulsive}(c) and (d), we mark the leading front of the hot channel by white arrows. The GOES light curves further suggest that the flare exhibits an extended maximum phase with dual peak structures at $\approx$18:00~UT and $\approx$18:13~UT. The comparison of AIA 94~\AA\ with co-temporal multi-channel RHESSI images clearly reveals two distinct regions of X-ray emission: the high energy HXR sources between 25 and 100 keV appear in pair as conjugate sources while the low energy emission below $\lessapprox$25 keV comes from the hotter region occupied with newly formed EUV coronal loops, in the wake of hot channel eruption (Figures~\ref{fig:RHESSI_AIA94_impulsive}(d)--(g)). We also find an increase in separation of the HXR footpoint sources as the flare progresses. We note that the strength of HXR emission to be higher at the southern footpoint of post-flare loop arcades where the HXR emission upto 50 keV was observed (Figure~\ref{fig:RHESSI_AIA94_impulsive}(g)), which suggests an asymmetry in the deposition of energy at conjugate footpoint locations. Soon after the second peak, the flaring region starts to show the formation of dense, bright, gradually rising post-flare loop arcades in the northern part (shown by red arrow in Figure \ref{fig:RHESSI_AIA94_impulsive}(g)). Gradually the southern part of the flaring region also exhibits the build-up of post-flare loop arcades (indicated by red arrow in Figure \ref{fig:RHESSI_AIA94_impulsive}(h)). In Figure \ref{fig:RHESSI_AIA94_impulsive}(i), we show well developed, dense post-flare loop arcades in both northern and southern part of the flaring region by red arrows. 

The comparison of evolution of HXR sources with respect to the AIA 304~\AA\ images are shown in Figure~\ref{fig:RHESSI_AIA304_impulsive}. After the activation of the filament, the brightening started to appear in the form of flare ribbons (shown by green arrows in Figure \ref{fig:RHESSI_AIA304_impulsive}(b)), which gradually move apart while exhibiting spatial expansion as well (cf. Figures~\ref{fig:RHESSI_AIA304_impulsive}(a)--(f)). Also, as expected, the high energy HXR sources of strength $\approx$25--100~keV show spatial consistency with the flare ribbons. The region marked by white arrow in Figure \ref{fig:RHESSI_AIA304_impulsive}(e) undergoes gradual increase in the brightness (cf. Figures \ref{fig:RHESSI_AIA304_impulsive}(e)--(i)). In the later stages, we observe distinct yet diffuse emission from post-flare coronal arcades, which is shown by black arrows in Figures \ref{fig:RHESSI_AIA304_impulsive}(g)--(i).

\begin{figure}
\epsscale{0.85}
\plotone{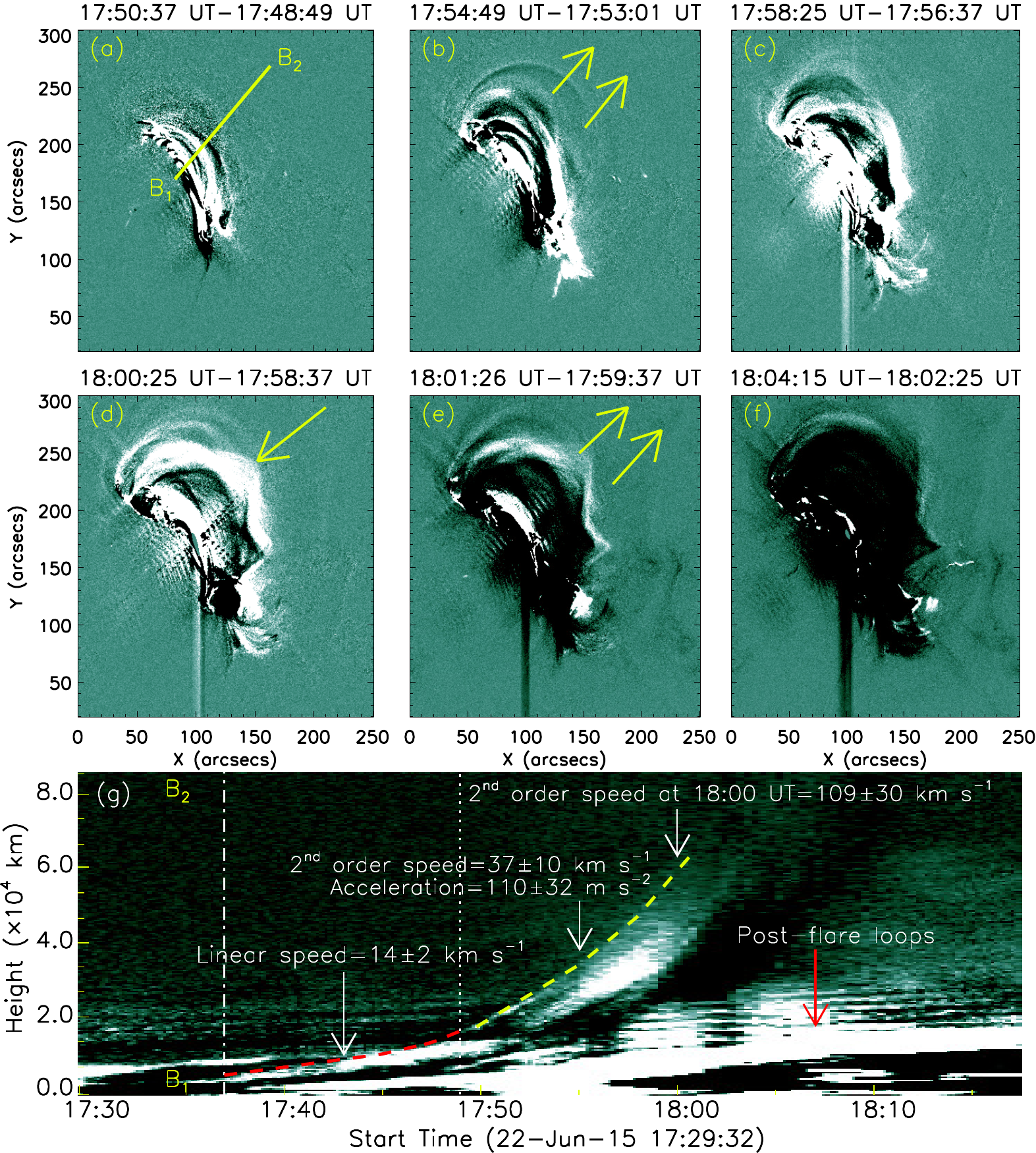}
\caption{Sequence of AIA 94 \AA\ running difference images showing the directions of eruption of the hot channel (shown by yellow arrows in panels (b) and (e)). The arrow in panel (d) shows the erupting front. We plot a time-slice diagram of the erupting hot channel in panel (g). The direction, from B$_{1}$ to B$_{2}$, through which the time-slice plot is drawn, is shown as a yellow slit in panel (a).}

\label{fig:AIA_diff}
\end{figure}

\begin{figure}

\epsscale{0.9}
\plotone{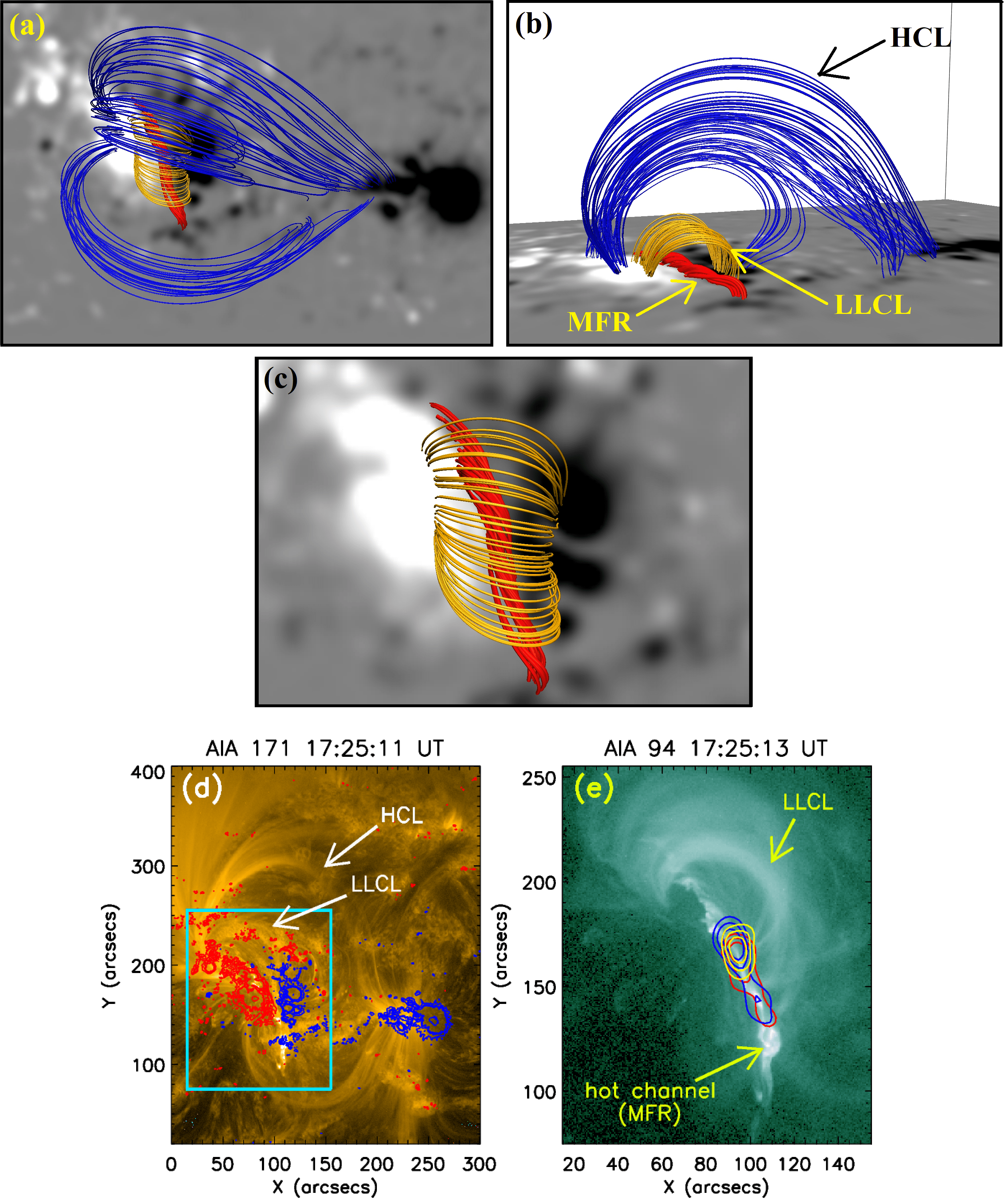}
\caption{Coronal magnetic field lines obtained using NLFFF model of extrapolation are shown in panels (a), (b), and (c). The lower boundary of the extrapolation is the photospheric LOS magnetic field. Panels (a) and (b) show the top and side view of the extrapolated field lines, respectively. The magnetic flux rope (MFR), low lying coronal loops (LLCLs), and high coronal loops (HCLs) are clearly indicated in panel (b). The position of the MFR along the PIL of the active region and the LLCLs are shown in panel (c). AIA 171 \AA\ image of the active region is shown in panel (d) in pre-flare phase (at $\approx$17:25 UT), overplotted with HMI LOS magnetogram. The positive and negative polarities of magnetogram are shown by red and blue contours respectively with contours levels set as $\pm$[400, 800, 1000, 2000] G. High and low coronal loops (HCL and LLCL, respectively) are shown by white arrows in panel (d). The rectangular box indicates a hot core region, whose enlarged view is shown in AIA 94 \AA\ channel in panel (e) overplotted with RHESSI contours in 5--10 keV (red), 10--15 keV (blue), and 15--25 keV (yellow). The X-ray contours are reconstructed by CLEAN algorithm with integration time of 40 seconds. The contours denote 70\%, 80\%, and 90\% of peak flux in each image. LLCLs are also clearly visible above the hot channel/MFR, which is co-spatial with X-ray sources.}
\label{fig:extrapolation_observations}
\end{figure}

As a comparison of phenomena occurring simultaneously in different heights of solar atmosphere, we have shown the flaring region in EUV (AIA 131 \AA) and UV (AIA 1600 \AA) channels during the first peak of the flare (Figure \ref{fig:AIA_131_1600}). AIA 131 \AA\ image clearly shows the erupting hot channel (i.e., MFR) structure (shown by white arrow in Figure \ref{fig:AIA_131_1600}(a)). We observe simultaneous conjugate and sheared flare ribbon brightenings at the photospheric level in AIA 1600 \AA\ image (marked by red arrows in Figure \ref{fig:AIA_131_1600}(b)).

To understand the activation and eruption of the hot channel, we plot a few AIA 94~\AA\ running difference images (Figure \ref{fig:AIA_diff}). In various panels, we mark the expanding hot channel by yellow arrows. As discussed earlier, the eruption resulted into a fast halo CME (Figure \ref{fig:cme}). In Figure~\ref{fig:AIA_diff}(g), we plot a time-slice diagram showing evolutionary phases of the hot channel. For the purpose, we have specified a narrow slit $\overline{B_{1}B_{2}}$, which is indicated in Figure~\ref{fig:AIA_diff}(a). The time-slice diagram is constructed using the running difference images with a time gap of $\approx$4 minutes between the successive images. The plot reveals a slow rise (speed $\approx$14 km~s$^{-1}$) phase of the hot channel between $\approx$17:37 UT and $\approx$17:49 UT (shown by red dashed line), which is followed by another phase of its fast eruption. A second order polynomial fit to the height-time measurement taken between $\approx$17:49 UT and $\approx$18:00 UT (shown by yellow dashed line) yields the speed of the erupting hot channel as $\approx$37~km~s$^{-1}$ with an acceleration of $\approx$110~m~s$^{-2}$. The speed of the erupting hot channel reaches to $\approx$109 km s$^{-1}$ at $\approx$18:00 UT. Notably, we observe formation of post-flare loops from $\approx$17:57~UT at a projected height of $\approx$4 Mm. 

In Figures \ref{fig:RHESSI_spectra}(d)--(h), we show RHESSI spectral fit results during the rise and main phase of the M6.6 flare. We find a steady rise in temperature as well as spectral hardening during the rise phase (Figure \ref{fig:RHESSI_spectra}(d)). In the main phase of the event (Figures \ref{fig:RHESSI_spectra}(e)--(h)) the spectra continued to become harder with $\delta\approx$3.3 at $\approx$18:24 UT. The maximum plasma temperature (T $\approx$27 MK) was observed around the second peak of the M6.6 flare (see Figure \ref{fig:RHESSI_spectra}(f)).

\subsection{Non-linear-force-free-field (NLFFF) modeling of active region corona}  
\label{sec:NLFFF_modeling}

The coronal magnetic field lines (Figures \ref{fig:extrapolation_observations}(a)--(c)) are extrapolated using the NLFFF model of \citet{Wiegelmann2008} to model the flux rope and associated coronal field lines. The lower boundary of the extrapolation is taken as photospheric LOS magnetogram. The MFR, low lying coronal loops (LLCLs), and high coronal loops (HCLs) are shown by arrows in Figure \ref{fig:extrapolation_observations}(b). Clear MFR is observed to form in between two opposite polarities of the trailing sunspot group (Figure \ref{fig:extrapolation_observations}(c)).

In Figures \ref{fig:extrapolation_observations}(d)--(e), AIA 171 \AA\ and AIA 94 \AA\ images of pre-flare stage (at $\approx$17:25 UT) distinctly show the hot channel (i.e., MFR), low lying coronal loops (LLCLs), and high coronal loops (HCLs). In Figure \ref{fig:extrapolation_observations}(d), we present 171 \AA\ image overplotted with photospheric LOS magnetogram with contour levels as $\pm$[400, 800, 1000, 2000] G. The blue and red contours denote negative and positive polarities, respectively. The rectangular box in Figure \ref{fig:extrapolation_observations}(d) shows a region of hot core, which contains the LLCLs and MFR. The enlarged view of rectangular box is shown in Figure \ref{fig:extrapolation_observations}(e) overplotted with RHESSI contours in 5--10 keV, 10--15 keV, and 15--25 keV. The contours denote 70\%, 80\%, and 90\% of peak flux in each image. Interestingly the X-ray contours are observed to be found along the length of the MFR.

\newpage

\section{Conclusions and discussions}
\label{sec:discussion}

In this paper, we provide a comprehensive multi-wavelength and multi-instrument study of a remarkable M-class major eruptive flare, which occurred in AR NOAA 12371 on 2015 June 22. The importance of the study lies in investigating the activities right from the early pre-flare phase till the decay of the flare with an aim of exploring the pre-flare processes in detail and the link between the pre-flare and main flare. The main observational results of the study are itemized below:

\begin{enumerate}

\item The eruption initiated from a magnetically bipolar region where a hot EUV channel (evidence of MFR) pre-existed (at least $\approx$5.5 hrs before the eruptive M6.6 flare) that exhibited early signatures of activation during the pre-flare activities. The H$\alpha$ observations reveal the presence of a filament in association with the coronal hot channel. Observations of the pre-flare phase clearly reveal activation of the filament with early eruption signatures, providing further credence of our interpretation of pre-flare activities. 

\item The hot channel is found to be co-spatial with a MFR detected in NLFFF model extrapolation. A very remarkable finding of the study lies in the detection of elongated as well as localized HXR sources of energies up to 25 keV that lie exactly over the extended central part of the hot channel. To our knowledge, this is the first time when an MFR has been detected in direct HXR observations. 

\item An important yet realistic coincidence is the continued presence of X-ray sources during the whole pre-flare phase. In the early pre-flare phase, the X-ray emission came from the core region, which were comprised of hot, dense bundle of low lying coronal loops, just above the filament channel. On the other hand, during the late pre-flare phase, as explained in item 2 above, the X-ray emission extended up to higher energies and the sources are located in the region where flux rope existed. These distinct pre-flare intensity enhancements, therefore, suggest build-up and activation of the MFR by magnetic reconnection involving interaction between the core field region and slowly evolving MFR.

\item The analysis of photospheric magnetograms during the extended period ($\approx$42 hours) prior to the pre-flare phase of the eruptive flare categorically reveals clockwise rotation of mix polarity sunspot group along with remarkable moving magnetic features.

\item With the onset of the impulsive phase of M6.6 flare, we find a sudden transition of the MFR from the state of slow rise ($\approx$14 km s$^{-1}$) to fast acceleration ($\approx$110 m s$^{-2}$ with the speed rises to $\approx$109 km s$^{-1}$ within AIA field-of-view), which points toward a feedback relationship between source region CME dynamics and the strength of the large-scale magnetic reconnection powering the eruptive flare. 

\item The classical signatures of large-scale magnetic reconnection are observed during the impulsive phase in terms of high energy (up to 100 keV) HXR conjugate sources that lie over the (E)UV flare ribbons. The H$\alpha$ observations show the remaining structures of the filament thus confirming the event to be a partial filament eruption.

\end{enumerate} 

The analyses carried out reveal the appearance of EUV hot channels in the corresponding SDO/AIA observations, well before ($\approx$5.5 hrs) the eruptive flare. This finding is in agreement with the contemporary understanding that the presence of MFR is a prerequisite for a CME \citep{Fan2005,Li2013,Song2019}. The correspondence between the spatial location of hot channel and MFR in coronal field modeling has been reported in several studies \citep{LiuTie2018,Mitra2018}, which is further confirmed by our analysis. However, the build-up mechanism of the MFR is still an open question, which requires extensive observational and theoretical research. The present study is a step in this direction and suggests that the pre-flare activities play an important role in the process of MFR activation. The pre-flare activity could be related to evolution in the photospheric magnetic structure. However, photospheric magnetic field changes in the active regions occur gradually but eventually lead to the development of complex magnetic field configuration in the corona which can also be seen in the present case. Our observation of rotation of sunspot group (in clock wise direction) over several hours, which encloses the PIL and, in the later phases, overlying developing MFR is probably related to the transfer of twist from sub-photospheric level to the coronal field lines. This long-lasting process would store excess magnetic energy into the coronal flux rope. The brightening up of the core field containing the MFR about 1.25 hours prior to the eruptive flare thus suggests the onset of heating, probably due to the magnetic reconnection, as the flux rope interacts with immediate low-lying arcades. Subsequently, the hot channel undergoes significant intensity enhancement and starts to appear in X-ray images up to 25 keV energies. Coronal pre-flare activity starts with the initiation of intense emission from the MFR and surrounding regions (an observational fact that has traditionally been observed in SXR as enhanced emission; see e.g., \citealt{Veronig2002,Chifor2007,Hernandez-Perez,Joshi2011,Joshi2013}).

We would like to highlight that, in our case, the regions of pre-flare activity and main M6.6 flare are co-spatial. The statistical studies carried out with SXR images from Yohkoh, revealed three categories of pre-flare activities in terms of source locations: co-spatial, adjacent, and remote \citep{Farnik1998,Kim2008}. The co-spatial and adjacent cases occurring within few minutes before the main flare are supposed to have direct relevance for the triggering processes related to the main flare \citep{Liu2009,Joshi2011,Mitra2019}. Notably, EUV and X-ray images clearly show that the pre-flare brightenings are spatially distributed along the hot channel (i.e., MFR) and within the core field region. Further, before the pre-flare emission, the region shows photospheric magnetic field changes along the PIL. These observations present consistency with the tether-cutting model of solar eruption \citep{Moore1992,Moore2001} where the build-up of MFR is a consequnce of flux changes along the PIL and, therefore, early reconnection signatures (causing the pre-flare activity) are expected to occur close to PIL and nearby core regions \citep{Yurchyshyn2006,Chen2014,Dhara2017}.

We find direct evidence of pre-flare tether-cutting reconnection in HXR imaging observations at lower energies (up to 25 keV). The HXR emission signifies intense heating of the core region. With the progress of pre-flare activity, the strength of HXR emission increases and a subtle yet clear non-thermal component starts to appear, which we identify as second pre-flare enhancement. The comparison of imaging and spectroscopic observations suggests that both thermal and non-thermal components originated from the EUV hot channel in the late pre-flare phase. Importantly, the HXR source at lower energies presents an elongated morphology and the X-ray sources lie exactly over the EUV hot channel. These observations provide direct support of tether-cutting reconnection. To our knowledge, the observation of extended HXR sources from a developing MFR during the pre-flare phase is a new observational finding. As expected, during this phase the hot channel rises slowly, an important feature of CME precursor \citep{Sterling2005,Nagashima2007,Sterling2007,Song2015}. In H$\alpha$ observations, the partial filament eruption begins at this time, further supporting the physical link between the pre-flare activity and initiation of solar eruption. Importantly, the slowly rising MFR transitioned to a phase of eruptive expansion with the onset of impulsive phase of M6.6 flare. Now, the appearance of `classical' flare signatures, viz., distinct coronal and footpoint HXR sources along with inner flare ribbons formed at both sides of PIL, provide evidence of large-scale magnetic reconnection, which are attributed to the reconnection-opening of overlying field lines (i.e., progression of reconnection in higher coronal fields of the envelope region) stretched by the erupting MFR. The sudden transition in the kinematic evolution of MFR from the phase of slow to fast rise precisely divides the pre-flare and impulsive phase of the flare, which we attribute to the feedback relationship between the early CME dynamics and the strength of the large-scale magnetic reconnection \citep{Temmer2008,Vrsnak2016,
Song2018,Mitra2019}.

\vspace{1.5em}
The authors would like to thank the SDO, RHESSI, and SOHO teams for their open data policy. SDO is NASA's mission under the Living With a Star (LWS) program. RHESSI was NASA's mission under the SMall EXplorer (SMEX) program. SOHO is a joint project of international cooperation between the ESA and NASA. This work is supported by the Indo-Austrian joint research project No. INT/AUSTRIA/BMWF/ P-05/2017 and OeAD project No. IN 03/2017. A.M.V. acknowledges the Austrian Science Find (FWF): P27292-N20. BBSO operation is supported by NJIT and US NSF AGS-1821294 grants. V.Y. acknowledges support from NSF AST-1614457, AFOSR FA9550-19-1-0040
and NASA 80NSSC17K0016, 80NSSC19K0257, and 80NSSC20K0025 grants. The authors sincerely thank the anonymous referee for providing useful suggestions and constructive comments, which enhanced the presentation and overall quality of the article.

\bibliographystyle{aasjournal}

\end{document}